\documentclass[longauth]{aa} 

\usepackage{graphicx}
\usepackage{amsmath}
\usepackage{subfigure}
\usepackage{natbib}
\bibpunct{(}{)}{;}{a}{}{,} 
\usepackage{txfonts}
\usepackage{epstopdf}
\usepackage{multirow}

\usepackage{amssymb}
\usepackage{amsmath}
\usepackage{mathtools}
\usepackage{algorithmic}
\usepackage{algorithm}
\usepackage{bm,bbm}

\DeclareMathOperator{\prox}{prox}

\newcommand{\cdbox}[1]{%
  \color{black}{.}%
  {\color{blue}%
    \dbox{\color{blue}#1}}%
}
\usepackage{dashbox,framed,color,ocg-p}
\newcommand{\ToggleLayer}[2]{%
  \leavevmode
  \pdfstartlink user {
    /Subtype /Link
    /Border [0 0 0]%
    /A <<
      /S/JavaScript
      /JS (
         var aOCGs = this.getOCGs(), Layer;
         var Layers = "#1".split(","), Active = -1, i, l;
         for (l=0; l<Layers.length; l++) {
           Layer = Layers[l];
           for (i=0; aOCGs && i<aOCGs.length; i++) {
             if (aOCGs[i].state && aOCGs[i].name == Layer) {
               Active = l;
               aOCGs[i].state = false;
             }
           }
           if (Active >= 0) break;
         }
         if (Active == -1) {
           for (l=0; l<Layers.length; l++) {
             if (Layers[l] == "") Active = l;
           }
         }
         Active = Active + 1;
         if (Active == Layers.length) Active = 0;
         Layer = Layers[Active];
         for (i=0; aOCGs && i<aOCGs.length; i++) {
           if (aOCGs[i].name == Layer) aOCGs[i].state = true;
         }
      )
    >>
  }#2%
  \pdfendlink
}

\begin{document}

\title{LOFAR sparse image reconstruction}

\author{
H.~Garsden\inst{1}\and 
J.~N.~Girard\inst{1}\and 
J.~L.~Starck\inst{1}\and 
S.~Corbel\inst{1}\and 
C.~Tasse\inst{2,3}\and 
A.~Woiselle\inst{4,1}\and 
J.~P.~McKean\inst{5}\and 
A.S.~van Amesfoort\inst{5}\and 
J.~Anderson\inst{6,7}\and 
I.~M.~Avruch\inst{8,9}\and 
R.~Beck\inst{10}\and 
M.~J.~Bentum\inst{5,11}\and 
P.~Best\inst{12}\and 
F.~Breitling\inst{7}\and 
J.~Broderick\inst{13}\and 
M.~Br\"uggen\inst{14}\and 
H.~R.~Butcher\inst{15}\and 
B.~Ciardi\inst{16}\and 
F.~de Gasperin\inst{14}\and 
E.~de Geus\inst{5,17}\and 
M.~de Vos\inst{5}\and 
S.~Duscha\inst{5}\and 
J.~Eisl\"offel\inst{18}\and 
D.~Engels\inst{19}\and 
H.~Falcke\inst{20,5}\and 
R.~A.~Fallows\inst{5}\and 
R.~Fender\inst{21}\and 
C.~Ferrari\inst{22}\and 
W.~Frieswijk\inst{5}\and 
M.~A.~Garrett\inst{5,23}\and 
J.~Grie\ss{}meier\inst{24,25}\and 
A.~W.~Gunst\inst{5}\and 
T.~E.~Hassall\inst{13,26}\and 
G.~Heald\inst{5}\and 
M.~Hoeft\inst{18}\and 
J.~H\"orandel\inst{20}\and 
A.~van der Horst\inst{27}\and 
E.~Juette\inst{28}\and 
A. ~Karastergiou\inst{21}\and 
V.~I.~Kondratiev\inst{5,29}\and 
M.~Kramer\inst{10,26}\and 
M.~Kuniyoshi\inst{10}\and 
G.~Kuper\inst{5}\and 
G.~Mann\inst{7}\and 
S.~Markoff\inst{27}\and 
R. McFadden\inst{5}\and 
D.~McKay-Bukowski\inst{30,31}\and 
D.~D.~Mulcahy\inst{10}\and 
H.~Munk\inst{5}\and 
M.~J.~Norden\inst{5}\and 
E.~Orru\inst{5}\and 
H.~Paas\inst{32}\and 
M.~Pandey-Pommier\inst{33}\and 
V.~N.~Pandey\inst{5}\and 
G.~Pietka\inst{21}\and 
R.~Pizzo\inst{5}\and 
A.~G.~Polatidis\inst{5}\and 
A.~Renting\inst{5}\and 
H.~R\"ottgering\inst{23}\and 
A.~ Rowlinson\inst{27}\and 
D.~Schwarz\inst{34}\and 
J.~Sluman\inst{5}\and 
O.~Smirnov\inst{2,3}\and 
B.~W.~Stappers\inst{26}\and 
M.~Steinmetz\inst{7}\and 
A.~Stewart\inst{21}\and 
J.~Swinbank\inst{27}\and 
M.~Tagger\inst{24}\and 
Y.~Tang\inst{5}\and 
C.~Tasse\inst{35}\and 
S.~Thoudam\inst{20}\and 
C.~Toribio\inst{5}\and 
R.~Vermeulen\inst{5}\and 
C.~Vocks\inst{7}\and 
R.~J.~van Weeren\inst{36}\and 
S.~J.~Wijnholds\inst{5}\and 
M.~W.~Wise\inst{5,27}\and 
O.~Wucknitz\inst{10}\and 
S.~Yatawatta\inst{5}\and 
P.~Zarka\inst{35}\and 
A.~Zensus\inst{10}
}

\institute{
Laboratoire AIM, Universit\'{e} Paris Diderot Paris 7/CNRS/CEA-Saclay, DSM/IRFU/SAp, 91191 Gif-sur-Yvette, France \and
Department of Physics and Electronics, Rhodes University, PO Box 94, Grahamstown 6140, South Africa \and
SKA South Africa, 3rd Floor, The Park, Park Road, Pinelands, 7405, South Africa \and
Sagem (Safran), 27 rue Leblanc, 75512 Paris Cedex 15, France \and
Netherlands Institute for Radio Astronomy (ASTRON), Postbus 2, 7990 AA Dwingeloo, The Netherlands \and
Helmholtz-Zentrum Potsdam, Deutsches GeoForschungsZentrum GFZ, Department 1: Geodesy and Remote Sensing, Telegrafenberg, A17, 14473 Potsdam, Germany \and
Leibniz-Institut f\"{u}r Astrophysik Potsdam (AIP), An der Sternwarte 16, 14482 Potsdam, Germany \and
SRON Netherlands Institute for Space Research, PO Box 800, 9700 AV Groningen, The Netherlands \and
Kapteyn Astronomical Institute, PO Box 800, 9700 AV Groningen, The Netherlands \and
Max-Planck-Institut f\"{u}r Radioastronomie, Auf dem H\"ugel 69, 53121 Bonn, Germany \and
University of Twente, The Netherlands \and
Institute for Astronomy, University of Edinburgh, Royal Observatory of Edinburgh, Blackford Hill, Edinburgh EH9 3HJ, UK \and
School of Physics and Astronomy, University of Southampton, Southampton, SO17 1BJ, UK \and
University of Hamburg, Gojenbergsweg 112, 21029 Hamburg, Germany \and
Research School of Astronomy and Astrophysics, Australian National University, Mt Stromlo Obs., via Cotter Road, Weston, A.C.T. 2611, Australia \and
Max Planck Institute for Astrophysics, Karl Schwarzschild Str. 1, 85741 Garching, Germany \and
SmarterVision BV, Oostersingel 5, 9401 JX Assen, The Netherlands\and
Th\"{u}ringer Landessternwarte, Sternwarte 5, D-07778 Tautenburg, Germany \and
Hamburger Sternwarte, Gojenbergsweg 112, D-21029 Hamburg, Germany \and
Department of Astrophysics/IMAPP, Radboud University Nijmegen, P.O. Box 9010, 6500 GL Nijmegen, The Netherlands \and
Astrophysics, University of Oxford, Denys Wilkinson Building, Keble Road, Oxford OX1 3RH, UK \and
Laboratoire Lagrange, UMR7293, Universit\'{e} de Nice Sophia-Antipolis, CNRS, Observatoire de la C\^{o}te d'Azur, 06300 Nice, France \and
Leiden Observatory, Leiden University, PO Box 9513, 2300 RA Leiden, The Netherlands \and
LPC2E - Universite d'Orl\'{e}ans/CNRS, France \and
Station de Radioastronomie de Nancay, Observatoire de Paris - CNRS/INSU, USR 704 - Univ. Orl\'{e}ans, OSUC , route de Souesmes, 18330 Nancay, France \and
Jodrell Bank Center for Astrophysics, School of Physics and Astronomy, The University of Manchester, Manchester M13 9PL, UK \and
Astronomical Institute ``Anton Pannekoek'', University of Amsterdam, Postbus 94249, 1090 GE Amsterdam, The Netherlands \and
Astronomisches Institut der Ruhr-Universit\"{a}t Bochum, Universitaetsstrasse 150, 44780 Bochum, Germany \and
Astro Space Center of the Lebedev Physical Institute, Profsoyuznaya str. 84/32, Moscow 117997, Russia \and
Sodankyl\"{a} Geophysical Observatory, University of Oulu, T\"{a}htel\"{a}ntie 62, 99600 Sodankyl\"{a}, Finland \and
STFC Rutherford Appleton Laboratory,  Harwell Science and Innovation Campus,  Didcot  OX11 0QX, UK \and
Center for Information Technology (CIT), University of Groningen, The Netherlands \and
Centre de Recherche Astrophysique de Lyon, Observatoire de Lyon, 9 av Charles Andr\'{e}, 69561 Saint Genis Laval Cedex, France \and
Fakult\"at f\"ur Physik, Universit\"at Bielefeld, Postfach 100131, D-33501, Bielefeld, Germany \and
LESIA, UMR CNRS 8109, Observatoire de Paris, 92195   Meudon, France \and
Harvard-Smithsonian Center for Astrophysics, 60 Garden Street, Cambridge, MA 02138, USA 
}

   \date{Received 30 June 2014; accepted 18 December 2014}

\abstract
{{\bf The LOFAR (LOw Frequency ARray) radio telescope is a giant digital phased array interferometer with multiple antennas distributed in Europe. It provides discrete sets of Fourier components of the sky brightness. Recovering the original brightness distribution with aperture synthesis forms an inverse problem that can be solved by various deconvolution and minimization methods.}}
{{\bf Recent papers have established a clear link between the discrete nature of radio interferometry measurement and the ``compressed sensing'' (CS) theory, which supports sparse reconstruction methods to form an image from the measured visibilities. Empowered by proximal theory, CS offers a sound framework for efficient global minimization and sparse data representation using fast algorithms. Combined with instrumental direction-dependent effects (DDE) in the scope of a real instrument, we developed and validated a new method based on this framework.}}
{{\bf  We implemented a sparse reconstruction} method in the standard LOFAR imaging tool and compared the photometric and resolution performance of this new imager with that of CLEAN-based methods (CLEAN and MS-CLEAN) with simulated and real LOFAR data.}
{{\bf We show that i) sparse reconstruction performs as well as CLEAN in recovering the flux of point sources, ii) performs much better on extended objects (the root mean square error is reduced by a factor of up to 10), and iii) provides a solution with an effective angular resolution 2-3 times better than the CLEAN images.}}
{{\bf  Sparse recovery gives a correct photometry on high dynamic and wide-field images and improved realistic structures of extended sources (of simulated and real LOFAR datasets). This sparse reconstruction method is compatible with modern interferometric imagers that handle DDE corrections (A- and W-projections) required for current and future instruments such as LOFAR and SKA.}}


\maketitle

\section{Introduction}
\label{intro}

Recent years have seen the development and planning of very large radio interferometers such as the LOw Frequency ARray (LOFAR) \citep{2013arXiv1305.3550V} in Europe, and the Square Kilometre Array (SKA) \citep{ska09} in Australia and South Africa (through its various precursors and pathfinders).

These new digital instruments have a very high sensitivity and a large field of view (FoV), as well as very large angular, temporal, and spectral resolutions in the radio spectrum observable from Earth. In particular, the very low frequency window (in the VHF - the very high frequency band between $\sim$10 and 250 MHz) is being explored (or revisited) with LOFAR, within the scope of various Key Science Projects, spanning the search for fast transients \citep{stappers2011A&A...530A..80S} to the study of early cosmology \citep{debruyn2012AAS...21921405D}.

\subsection{LOFAR}
LOFAR is a digital radio interferometer composed of 48 stations: 40 stations constitute the core and remote stations (two of which are future stations) that are distributed in the Netherlands (NL), and eight international stations that are located in Germany, Great Britain, France, and Sweden. Poland has planned the construction of three international stations that will enrich the array in the east-west direction. LOFAR has two working bands: the LBA-band (low-band antenna) from $\geq$30 to 80 MHz (that can be switched to 10-80 MHz), and the HBA-band (high-band antenna) from 110 to 250 MHz, lying on either-side of the FM band. One station consists of two arrays composed of fixed crossed dipole antennas, each offering a large FoV (therefore low directivity) and broadband properties\footnote{see details at www.lofar.org}. The HBA field of the core stations is split into two to increase the number of baselines.The stations measure induced electric signals that undergo pre-processing operations in the station back-end, consisting of digitization, filtering, phasing, and summing. All these steps constitute the beamforming step of the phased antenna array. The output signal of one station is thus similar to that of a synthetic antenna whose beam is electronically pointed (rather than mechanically) in the direction of interest. Since most steps are performed in the digital side, multi-beam observations are possible and are only limited by the electronic hardware (i.e. in field-programmable gate arrays -- FPGAs, by making trade-offs between the observed bandwidth and the number of numerical beams pointing at the sky).

At the interferometer level, the signal from every station is combined in a central correlator in the NL that performs a phased sum or inter-station cross-correlation. The latter operation enables LOFAR to build up a digital interferometer that samples a wide range of baselines (from $\sim$70 m up to $\sim$1300 km). The high flexibility of the instrument comes with the necessity of using advanced calibration strategies depending on the type of observations and its expected final sensitivity.

Since late 2012, LOFAR has been opened to the astronomical community by a regular public call for proposals\footnote{see the observation proposal section on www.astron.nl}.
Early results with LOFAR demonstrated that it can reach a very high dynamic range (DR) on a wide FoV (i.e. several tens of degrees, see for example \citet{yatawatta2013A&A...550A.136Y}) and a very high angular resolution \citep{wucknitz2010,shulevski} at low frequencies. 

LOFAR capabilities are geared to key science projects on deep extragalactic surveys, transient radio phenomena and pulsars, the epoch of reionization, high-energy cosmic rays, cosmic magnetism, and solar physics and space weather.

\subsection{Increased complexity of low-frequency imaging}
Since the beginning of radio interferometry imaging, various imaging methods have been designed to fit the requirements of different types of (extended) radio objects. The availability of high-performance computing and the need for efficient, fast, and accurate imaging for new wide-field interferometers has motivated the implementation of new imaging algorithms. Given the recording time and frequency resolutions, the integration time, and the diversity of baselines of wide-field interferometers, large amounts of data storage are required to save the telescope data. These data must then be transformed into a scientifically exploitable form (typically into images cubes). Substantial computational power is required for this as well \citep{begeman11}.

Because of the nature and the dimensions of the LOFAR array, direction-dependent effects (DDE) \citep{tasse2012} occur during the span of a LOFAR observation and add up to the usual other effects intervening in classical radio interferometers. These effects require a direction-dependent calibration before imaging. In particular, the classical compact planar array and small FoV assumptions are no longer valid, especially for a wide-field instrument such as LOFAR. Modern calibration and imaging at radio wavelengths require a similar approach to that used in modern optical telescope with adaptive optics.

The problem can be generalized and expressed in the Measurement Equation framework (Sect. \ref{DDE}). The calibration problem therefore manifests as an inverse problem that should be solved to independently determine all the parameters and coefficients that describe each observation dataset.

Among the recent developments of data processing and reconstruction algorithms, the discovery of compressed sensing (CS) \citep{candes06} has led to new approaches to solving these problems. It has been proposed for radio interferometry (e.g. \citet{wiaux09,Wenger2010,Li2011RMS,Li2011,carrillo12}) because this constitutes a relevant practical case as a result of the sparse nature of the interferometric sampling (Sect. \ref{descCS}).

The implementation of sparse radio image reconstruction methods is expected to produce better results on large extended objects with high angular resolution than other classical deconvolution methods. In Sect. \ref{imaging} we first apply the theory of sparse reconstruction within the scope of radio aperture synthesis imaging and then implement it in the LOFAR imaging software. We also relate it to previous CLEAN-based deconvolution methods. We present in Sect.  \ref{benchmark} the results of benchmark tests using simulated and real LOFAR datasets by focusing on the quality of the image reconstruction compared to that of usual CLEAN-based algorithms. We then discuss the practical advantages and limitations of the current implementation and possible future developments.



\section{Image reconstruction: from CLEAN to compressed sensing}
\label{imaging}
\subsection{Introduction}

An ideal radio interferometer, composed of co-planar and identical antennas, samples the sky in the Fourier domain \citep{wilson09}. In other words, each pair of antennas that forms one baseline gives access to the measure of the sky brightness as seen through a set of fringes with  characteristics that depend on the frequency, the baseline length, and orientation with respect to the source.
As for optical interferometers, the measured quantity (after correlating the pair of signals) is the fringe contrast, also called (fringe) visibility. In the radio domain where the electric field varies slowly, we can sample both phase and amplitude with time and frequency. In the scope of radio interferometry, the visibility function is a complex function.

An N-antenna ideal interferometer provides $N(N-1)/2$ independent instantaneous visibility measurements. We define the spatial frequency coordinates plane, the (u,v) plane, which is orthogonal to the direction of observation containing all projected baselines (the third coordinate, w, is omitted in the small-field approximation). This direction defines the origin of the Fourier conjugated coordinate system on the sky, the direction cosines (l,m).

\begin{figure}
\begin{center}
\includegraphics[width=9cm]{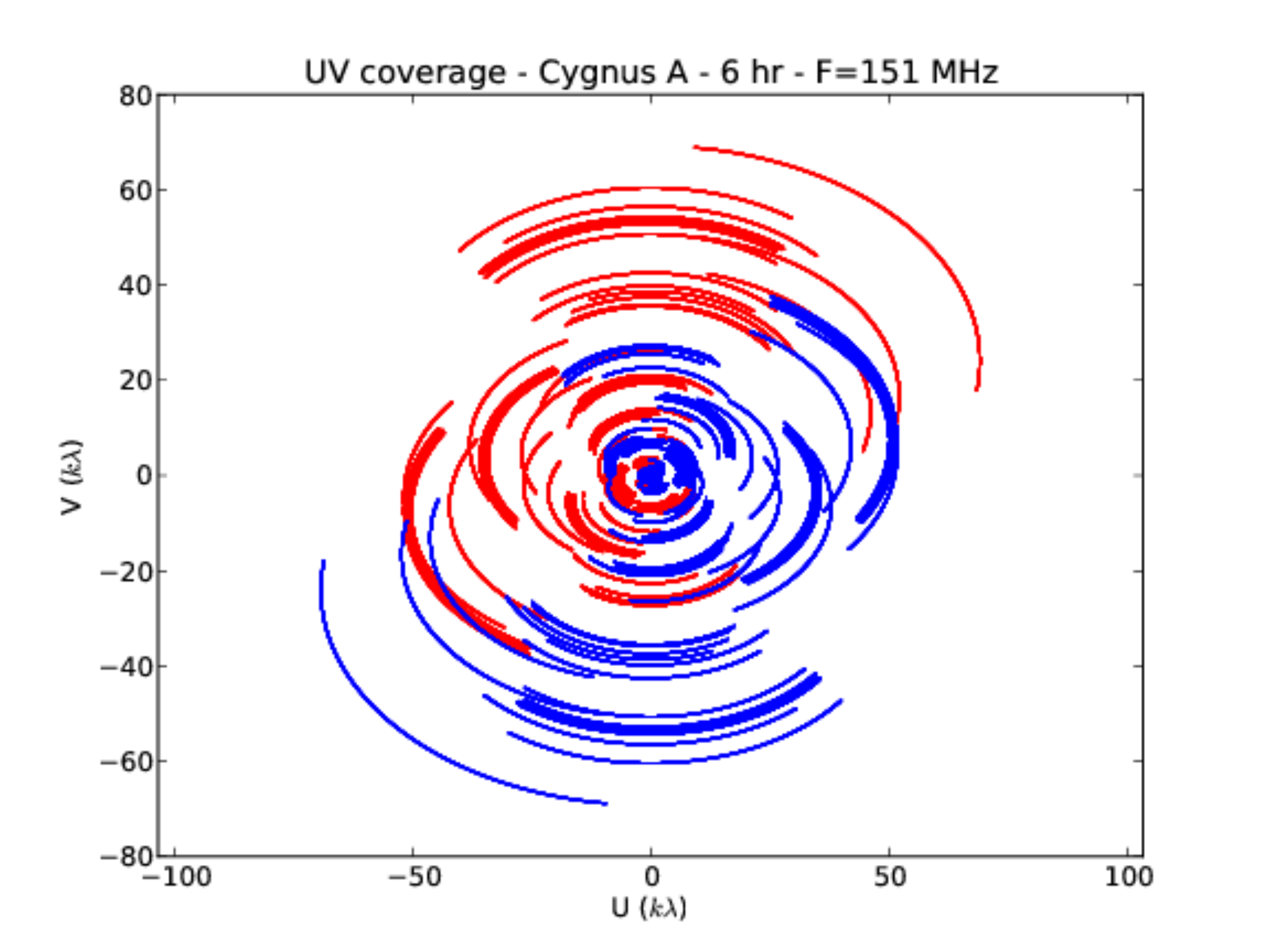}
\caption{Visibility coverage from a six-hour observation of Cygnus A \citep{mckean10}. Visibilities are plotted in the Fourier space with the U (x-axis) and V (y-axis) being the spatial frequencies in wavelength unit, $\lambda$ (here f=151 MHz, $\lambda\approx$ 2 m) determined by the baseline projection on the sky. Each (u,v) point (red) has its symmetric (-u,-v) point (blue) corresponding to the same baseline. The lines indicate (u,v) points where a visibility was recorded. The arcs are built from varying the baseline projections with the rotation of Earth during the observation.}
\label{visibilities}
\end{center}
\end{figure}

At any given time and frequency, one projected baseline samples one spatial frequency represented by one point in the (u,v) plane. With time, each (u,v) point sweeps the (u,v) plane, which enriches it with new samples (i.e. \textit{Earth rotation synthesis}). Figure \ref{visibilities} presents the (u,v) coverage of one typical LOFAR observation integrated over six hours. Each track is associated with one antenna pair, and its shape depends on the antenna configuration, the instrument latitude, and the direction of observation.

The number of visibilities tends to increase with the building of million-element interferometers such as SKA, but stays small in case of snapshot (short integration time) observations. This collection of Fourier samples is the starting point for the imaging process, which consists of using the measurements to recover the true visibility function in the Fourier domain.
After instrumental calibration, a dirty image can be generated by gridding the calibrated visibilities to a 2D plane by inverting this approximate image of the true Fourier transform to the image plane. The missing samples contribute to corrupt the image because it is a rough approximation of the sky brightness. This image is the convolution of the true sky brightness (represented as a 2D image) with the interferometer point spread function (PSF, or dirty beam).

We now discuss the different approaches that can be used to approach the true sky brightness from the measured visibilities, starting with the CLEAN method, before describing the method based on the sparsity of the measured signal.

\subsection{CLEAN}
For many years, deconvolution has been achieved through the CLEAN algorithm \citep{hogbom74} and its variants \citep{clark80,schwab84}.
CLEAN considers the dirty image to be constructed from point sources convolved with the PSF; extended objects are decomposed as point sources as closely as possible. CLEAN operates directly on the dirty image (original versions like H\"ogbom do) by locating the maximum of the image and iteratively subtracting a fraction of the dirty beam centred and scaled to the located maximum. The detected sources are indexed as CLEAN components to form a model image enriched by the successive subtraction steps.
The source detection and subtraction continues until a threshold is reached on the residual image (typically representing the background level), which is assumed to contain no remaining sources. The model image contains a pixel-wise description of the levels and locations of the detected sources.
To remove the unphysically high spatial frequency components (associated with the pixel size) that are introduced by the CLEAN algorithm, the model image is usually convolved by an elliptical Gaussian 2D fit of the centre of the dirty beam (which is the CLEAN beam) to provide a final angular resolution corresponding to that accessible by the interferometer. The residual image, containing no other source, is added to the convolved CLEAN image to form the restored image and represents the noisy sky background.

This image can be described in the following manner:
\begin{equation}
\mathcal{I} = \mathcal{M} \ast C + \mathcal{R}
\label{clean image}
,\end{equation}
where $\mathcal{I}$ is the restored image, $\mathcal{M}$ is the model composed of CLEAN components, $C$ is the CLEAN beam, $\mathcal{R}$ is the residual, and $\ast$ is the convolution operator.

The CLEAN method has several variants, one of which is Cotton-Schwab CLEAN (CoSch-CLEAN) from \citet{schwab84}, which was used for all experiments in this article. 
CLEAN can be described as a least-squares minimization that solves normal equations and its variants that are usually split into \textit{major} and \textit{minor} cycles by combining a steepest-descent method (major cycle) and a greedy algorithm (minor cycle), which quickly calculates an approximate solution of the normal equations.

CoSch-CLEAN uses the steepest-descent method (during the major cycle) combined with a greedy algorithm (during the minor cycle) that calculates an approximate solution of the normal equations.


With LOFAR, the PSF varies over the FoV and is also difficult to determine accurately because of instrumental DDEs (Sect. \ref{DDE}). Moreover, the sensitivity and variety of baseline \textbf{s} brought by the instrument often limit the quality of the restored images using these methods.

\subsection{Toward multi-resolution}
The CLEAN method is well known to produce poor solutions when the image contains large-scale structures. \citet{rest:wakker88} introduced the concept of multi-resolution CLEAN (M-CLEAN) to alleviate difficulties occurring in CLEAN for extended sources. The M-CLEAN approach consists of building two intermediate images, the first one (called the smooth map) by smoothing the data to a lower resolution with a Gaussian function, and the second one (called the difference map) by subtracting the smoothed image from the original data.  Both images are then processed separately. By using a standard CLEAN algorithm on them, the smoothed clean map and difference clean map are obtained. The recombination of these two maps gives the clean map at the full resolution.

This algorithm may be viewed as an artificial recipe, but it has been shown that it is linked to wavelet analysis \citep{aper:starck91}, leading to a wavelet CLEAN (W-CLEAN) method \citep{starck:sta94_3}.
Furthermore, in the W-CLEAN algorithm, the final solution was derived using a least-squares iterative reconstruction algorithm applied to the set of wavelet coefficients detected by applying CLEAN on different wavelet scales. This can be interpreted as a debiased post-processing of the peak amplitudes found at the different scales. A positivity constraint on the wavelet coefficients was imposed during this iterative scheme by nullifying the negative coefficients.
 
Other multi-scale approaches exist, such as the adaptive scale pixel (Asp) deconvolution \citep{asp2004} (which fits for parameters of Gaussian sets), the multi-scale CLEAN (MS-CLEAN) \citep{Cornwell:msclean2008}, and the multi-scale multi-frequency (MS-MF) deconvolution \citep{rau11}.
On the one hand, Asp is not implemented in the LOFAR imager, and on the other hand, we are imaging single-frequency channel datasets that do not justify the use of MS-MF even though they are in the LOFAR imager. We therefore only use MS-CLEAN in this paper as a scale-sensitive algorithm. It consists of fitting a collection of extended patches (or blobs). Instead of subtracting the dirty beam at a given location of the residuals at each iteration, as in the CLEAN algorithm, MS-CLEAN subtracts a blob, estimating first the most adequate blob size. MS-CLEAN presents several problems that do not exist in M-CLEAN or W-CLEAN, however. Indeed, it is unclear which function the algorithm minimizes, or even if it minimizes anything at all. The varying background is also problematic in MS-CLEAN (as in CLEAN), while it is not in W-CLEAN. MS-CLEAN also relies on an arbitrary manual setting of the characteristic scales of the image.

However, these different CLEAN-based algorithms (hereafter denoted x-CLEAN) have all shown to significantly improve the results over CLEAN when the image contains extended sources.

Other methods, based on a statistical Bayesian approach, have been developed to image extended emission \citep{junklewitz13,sutter14} but were not included in the present work. 

\subsection{Compressed sensing and sparse recovery}
\label{descCS}
Compressed sensing (CS) \citep{candes06,donoho06} is a sampling and compression theory based on the sparsity of an observed signal, which shows that under certain conditions, one can exactly recover a $k$-sparse signal (a signal for which only $k$ coefficients have values different from zero out of $n$ total coefficients, where $k < n$) from $m < n$ measurements. CS requires the data to be acquired through a random acquisition system, which is not the case in general. However, even if the CS theorem does not apply from a rigorous mathematical point of view, some links can still be considered between real life applications and CS. 
In astronomy, CS has already been studied in many applications such as data transfer from satellites to Earth \citep{bobin,barbey2011}, 3D weak lensing \citep{leonard2012}, next-generation spectroscopic instrument design \citep{Ramos11}, and aperture synthesis. Indeed, there is a close relationship between CS principles and the aperture-synthesis image reconstruction problem, which was first addressed in \citet{wiaux09}, \citet{Wenger2010}, \citet{Li2011} and \citet{carrillo12}. Wide-field observations were subsequently studied in \citet{McEwen2011}, and different antenna configurations \citep{Fannjiang2013} and non-coplanar effects \citep{wiaux09,wiauxspread09,wolz2013} were analysed in a compressed-sensing framework.
Aperture synthesis presents the three main ingredients that are fundamental in CS: 
 \begin{itemize}
\item \textbf{Underdetermined problem:} we have fewer measurements (i.e. visibilities) than unknowns (i.e. pixel values of the reconstructed image).
\item \textbf{Sparsity} of the signal: the signal to reconstruct can be represented with a small number of non-zero coefficients. For point source observations, the solution is even strictly sparse (in the Dirac domain) since it is only composed of a list of spikes. For extended objects, sparsity can be obtained in another representation space such as wavelets.
\item \textbf{Incoherence} between the acquisition domain (i.e. Fourier space) and the sparsity domain (e.g. wavelet space). Point sources, for instance, are localized in the pixel domain, but spread over a large domain of the visibility plane. Conversely, each visibility contains information about all sources in the FoV.
\end{itemize}

From the CS perspective, the best way to reconstruct an image $X$ from its visibilities is to use sparse recovery by solving the following optimization problem:
\begin{equation}
  \min_{}  \Arrowvert X \Arrowvert_{p}\ \ \text{ subject to } \ \ \Arrowvert V - \mathbf{A} X \Arrowvert_{2}^{2} < \epsilon,
\label{minimize_l0_pix}
\end{equation}
where $V$ is the measured visibility vector, $\mathbf{A}$ the operator that embodies the realistic acquisition of the sky brightness components (instrumental effects, DDE, etc.), and   $\| z \|^p_p=\sum_i |z_i|^p$. To reinforce the sparsity of the solution, $p$ is positive but must be lower than or equal to 1.
In particular, for $p=0$, we derive the $\ell_0$ pseudo-norm that counts the number of non-zero entries in $z$. 
The first part of Eq. \ref{minimize_l0_pix} enforces sparsity, the second part indicates that the reconstruction matches the visibilities within some error $\epsilon$.
Most natural signals, however, are not exactly sparse but rather concentrated within a small set.
Such signals are termed \textit{compressible} or \textit{weakly sparse}, in the sense that the sorted coefficient values decay quickly according to a power law. The faster the amplitudes of coefficients decay, the more compressible the signal is.
   
It is interesting to note that the CLEAN algorithm can be interpreted as a \textit{matching-pursuit algorithm} \citep{lannes1997} that minimizes the $\ell_0$ pseudo-norm of the sparse recovery problem of Eq.~\ref{minimize_l0_pix}, but recent progress in the field of numerical optimization (see application with the SARA algorithm in \citet{Carrillo2014} and references therein) allows us to have much faster algorithms. 
 The sparsity model is extremely accurate for a field containing point sources, since the true sky can in this case be represented just by a list of spikes.
 This explains the very good performance of H\"ogbom's CLEAN for point sources recovery, and why astronomers still use it 40 years after it has been published.
But CS provides a context in which we can understand the limitation of CLEAN for extended object reconstructions. The notion of sparsity in radio astronomy is not new and may be recognized in algorithms that apply a transform to the data. For example, extended objects are not sparse in the pixel basis, therefore sparse recovery algorithms cannot provide good solutions on this basis. Indeed, as depicted in \citet{rest:wakker88} and \citet{starck:sta94_3} with wavelets, representing the data in another domain where the solution is sparse was shown to be a good approach.
 More generally, we can assume that the solution $X$ can be represented as the linear expansion 
\begin{equation}
X =\mathbf{\Phi}\alpha=\sum_{i=1}^{t}\phi_{i}\alpha_i ~,
\end{equation}
where $\alpha$ are the synthesis coefficients of $X$, $\mathbf{\Phi}=(\phi_{1},\ldots,\phi_{t})$ is the dictionary whose columns are $t$ elementary waveforms $\phi_{i}$ also called \textit{atoms}.  The dictionary $\mathbf{\Phi}$ is a $b\times t$ matrix whose columns are the normalized atoms (of size $b$), 
 assumed here to be normalized to a unit $\ell_{2}$-norm, 
 that is,  $\forall i\in[1,t],\left\Vert \phi_{i}\right\Vert _2^2=\sum_{n=1}^{N}\left|\phi_{i}[n]\right|^{2}=1$.
The minimization problem of Eq.~\ref{minimize_l0_pix} can now be reformulated in two ways, the synthesis framework
\begin{equation}
  \min_{\alpha}  \Arrowvert \alpha \Arrowvert_{p}\ \ \text{ subject to } \ \ \Arrowvert V - \mathbf{A}\mathbf{\Phi} \alpha  \Arrowvert_{2}^{2} < \epsilon,
\label{minimize_lp_synthesis}
\end{equation}
and the analysis framework
\begin{equation}
  \min_{\alpha}  \Arrowvert \mathbf{\Phi}^t X \Arrowvert_{p}\ \ \text{ subject to } \ \ \Arrowvert V - \mathbf{A} X \Arrowvert_{2}^{2} < \epsilon.
\label{minimize_lp_analysis}
\end{equation}
When the matrix $\mathbf{\Phi}$ is orthogonal, both analysis and synthesis frameworks lead to the same solution, but the choice of the synthesis framework is generally the best if $X$ has a clear and unambiguous representation in $\mathbf{\Phi}$.

In that scope, the recent proximal theory can provide proof of convergence of algorithms constructed in this framework (in
contrast to some CLEAN derivates). It also led to the development of fast optimization algorithms combined with fast transforms enabling a fast (usually $N \log N$) evaluation of atom coefficients.
The best dictionary provides the sparsest (i.e. the most economical) representation of the signal. 
In practice, implicit dictionaries with fast transforms (such as the wavelet or curvelet transforms, etc.) allow us to directly obtain the coefficients and reconstruct the signal using fast algorithms running in linear or almost linear time (unlike matrix-vector multiplications).




\subsection{Imaging with LOFAR}
\label{DDE}
For a large multi-element digital interferometer such as LOFAR observing in a wide FoV, the small-field approximation is no longer valid, and the sampled data (represented by the operator $\mathbf{A}$) are no longer a discrete set of 2D Fourier components of the sky. The instrumental (direction-independent) and natural direction-dependent effects have to be taken into account for proper image reconstruction.
These effects include
\begin{itemize}
\item the instrumental effects such as inter-station clock shifts,
\item the non-coplanar nature of the baselines \citep{cornwell1992} and the effect of their projections (the sample visibility function has a non-zero third coordinate, i.e. $w\neq0$), 
\item the anisotropic directivity of the phased-array beam \citep{bhatnagar08}, and the non-trivial dipole projection effects with time and frequency,
\item the sparsity in the sampling of the visibility function (i.e. the limited number of baselines and the time/freq integration),
and \item the effect of the interstellar- and interplanetary media, and Earth's ionosphere, on the incoming plane waves.
\end{itemize}

These effects can be modelled in the framework of the radio interferometry measurement equation (RIME, see \citet{hamaker96p1,hamaker96p2,smirnov2011} and following papers), which describes the relation between the sky and the four-polarization visibilities associated with each pair of antennas in a time-frequency bin. It can model all the DDE mentioned above, cumulated in 2$\times$2 \textit{Jones} matrices that influence the electric field and voltage measurements. One of the most basic ways to express a four-polarization visibility of one baseline using the RIME formalism is as follows (in a given time-frequency bin) :

\begin{equation}
V^{\text{real}}_{pq} = \mathbf{J_p}^{-1}V^{\text{mes}}_{pq}(\mathbf{J^H_q})^{-1}
\label{eqJones}
,\end{equation}
with
\begin{itemize}
\item[$V^{\text{mes}}$] the four-polarization (XX,XY,YX,YY) measurement matrix corrupted by the DDE,
\item[$V^{\text{true}}$] the true visibility matrix uncorrupted by the DDE,
\item[$\mathbf{J_p}$] the cumulated \textit{Jones} matrix of antenna p, and
\item[$\mathbf{J_q^{H}}$] the Hermitian transpose (conjugated transpose) of the cumulated \textit{Jones} matrix of antenna q.
\end{itemize}

The corrected visibility $V^{\text{corr}}_{pq}$ matrix can be expressed as a four-dimensional vector (which also depends on the time $t$ and frequency $\nu$) as in Eq. 1 of \citet{tasse13},
\begin{equation}
\text{Vec}(V^{\text{corr}}_{pq})=\int_S (\mathbf{D^{t\nu,*}_{q,\vec{s}}} \otimes \mathbf{D^{t\nu}_{p,\vec{s}}})\text{Vec}(I_{\vec{s}})\times e^{-2i\pi\phi(u,v,w,\vec{s})} d\vec{s}
\label{VectDDE}
,\end{equation}
where $I_s$ is the four-polarization sky, \vec{s} the pointing direction, $\otimes$ the Kronecker product producing a 4$\times$4 matrix (referred to as the Mueller matrix),
$\text{Vec}()$ the operator that transforms a 2$\times$2 matrix into a four-dimensional vector, $\mathbf{D^{t\nu,*}_{q,\vec{s}}}\otimes \mathbf{D^{t\nu}_{p,\vec{s}}}$, the 4$\times$4 Mueller matrix, containing the accumulation of the direction-dependent terms and the array geometry (see details in Appendix A in \citet{tasse13}), and $\phi(u,v,w,\vec{s})=ul+vm+w(\sqrt{1-l^2-m^2}-1)$, the baseline and direction-dependent phase factor.
Because the sky is described in terms of Stokes intensity (I, Q, U, V), and not in terms of the electric field value, the RIME is linear in its sky-term. By considering the total set of visibilities over baselines, time, and frequency bins, and by including the measurement noise $\epsilon$, we can therefore reduce Eq. \ref{VectDDE} to
\begin{equation}
V=\mathbf{A}I_s+\epsilon
\label{eqVis}
,\end{equation}
with
\begin{itemize}
\item[$I_s$] the four-polarization sky,
\item[$\mathbf{A}$] the transformation matrix from the sky to the visibilities, including all DDE, and
\item[$\epsilon$] the measurement noise.
\end{itemize}

The structure of matrix $\mathbf{A}$ and its connection to the RIME and Jones formalism shown above is described in detail in \cite{tasse2012}.
The linear operator $\mathbf{A}$ contains (i) the Fourier kernels and the information on (ii) the time-frequency-baseline dependence of the effective 2$\times$2 Jones matrices, and (iii) the array configuration that is represented by the (u,v,w)-sampling over time and frequency. It is important to note that both (ii) and (iii) cause $\mathbf{A}$ to be non-unitary, which is a very important fact for the work presented in this paper. 
In practice, it is virtually impossible to make $\mathbf{A}$ explicit, essentially because of its $N_{vis} \times N_{pixels}$ dimension. Instead, $\mathbf{A}$ and $\mathbf{A^T}$ can be applied to any sky or visibility vector respectively using {\em A-projection} \citep{bhatnagar08,tasse13}. To cope with the non-coplanar effect, the {\em W-projection} algorithm \citep{cornwell08} is used to turn the 3D recorded visibilities into a 2D Fourier transform.
In the scope of the LOFAR project and its data reduction pipeline, the \textit{AWimager} \citep{tasse13} program was developed for imaging the LOFAR data by taking into account both A- and W-projections. It is therefore of high interest to implement CS in this type of new-generation imagers.

\subsection{Algorithm}
\label{AlgoMAD}
To perform the sparse image reconstruction, we need to
\begin{enumerate}
\item  select a minimization method to solve Eq.~\ref{minimize_lp_synthesis} or Eq.~\ref{minimize_lp_analysis}, 
\item  select a dictionary $\mathbf{\Phi}$, and 
\item  select the parameter related to the minimization method.
\end{enumerate}

Several minimization methods have been used for aperture synthesis, 
the FISTA method (fast iterative shrinkage-thresholding algorithm in \citet{BeckFista09}) in \citet{Li2011} ,\citet{Wenger2010}, \citet{hardy13} and \citet{Wenger2013},  
the OMP (orthogonal matching pursuit) \citep{DMA97} in \citet{Fannjiang2013}, Douglas-Rachford splitting in \citet{wiauxspread09}, \citet{McEwen2011} and \citet{carrillo12} or the SDMM (simultaneous-direction method of multipliers in \citet{CombettesPesquet09} and \citet{Carrillo2014}.
In our experiments, we investigated mainly 
two algorithms, FISTA, 
and a recent algorithm proposed by \citet{vu2013splitting}, which works in the analysis framework. As they were providing very similar results, we chose to report here results derived with the FISTA 
algorithm. Full details can be found in  \citet{BeckFista09}, \citet{Wenger2010}, and \citet{starck:book10}.
 
The choice of dictionary is critical. Optimal dictionaries should contain atoms that represent  the content of the data well.
 In this sense, the starlet transform, also called isotropic undecimated wavelet transform \citep{starck:book10}, which  was used in \citet{Li2011}, is a very good choice, since this decomposition has shown to be extremely useful for astronomical image restoration \citep{starck:book06}.
 In contrast, orthogonal or bi-orthogonal wavelets, even using several decompositions, are well known to produce artefacts due to critical sampling. 
 Other dictionaries such as curvelets \citep{starck:book10} are a good alternative if the data contain directional features, such as jets, which will be poorly represented with wavelets. The block discrete cosine transform (BDCT) used in \citet{Fannjiang2013} is relatively hard to justify for astronomical images, since its atoms present an oscillatory pattern. 
In our study, we focus on the starlet dictionary.

The convex optimization problem formulated in Eq. \ref{minimize_lp_synthesis} can be recast in an augmented Lagrangian form with
\begin{equation}
\min_{\alpha}  \Arrowvert V - \mathbf{A}\mathbf{\Phi} \alpha \Arrowvert_{2}^{2} + \sum_j \lambda_j | \alpha_j | 
\label{minimize_l0_pix_2}
.\end{equation}

The first part of Eq. \ref{minimize_l0_pix_2} indicates that the reconstruction should match the data and the second term enforces sparsity (through controlling the regularization parameter $\lambda_j$ at each scale $j$ of $\mathbf{\Phi}$). The information about the error $\epsilon$ in Eq. \ref{minimize_l0_pix} is included in $\lambda_j$.
The problem can then be solved using the FISTA algorithm \citep{BeckFista09}, following the implementation of algorithm \ref{Fista}.


The final problem remains: choosing parameters that are needed to control the algorithm. Most minimization methods, using $l_0$ and $l_1$, have a single thresholding step, where coefficients in the dictionary have to be soft or hard-thresholded using a threshold value, which is a value $\lambda$, common to projection in $\Phi$ (unlike Eq. \ref{minimize_l0_pix_2}) (see \citet{starck:book10} for more details on hard and soft thresholding).
This parameter controls the trade-off between the fidelity to the observed visibility and the sparsity of the reconstructed solution.
In \citet{Li2011} and \citet{Wenger2013}, $\lambda$ was fixed to an arbitrary value, different for each experiment, and certainly after several tests. This approach may be problematic for real data where the true solution is not known. 

\citet{Carrillo2014} addressed the constrained problem of Eq.~\ref{minimize_lp_analysis} directly and the $\epsilon$ parameter is a bound on the $\ell_2$-norm data fidelity term, which is related explicitly to the noise level under the assumption that the noise is \textit{i.i.d.} Gaussian. As the noise is generally not white, a whitening operator has to be applied first.

We propose in this paper another strategy, where the threshold is fixed only from the noise distribution at each wavelet scale $j$. Indeed, estimating a threshold from the residuals in a pixel basis is far from being optimal. On the one hand, the residual is not necessarily of zero mean and the background level is not known, and, on the other hand, the signal can be at the noise level in the pixel basis, but much stronger in the wavelet basis.

This has two main advantages:  the default threshold value should always give reasonable results, and it optimizes the probability detection of faint objects. Indeed, an arbitrary threshold value could lead to many false detections in the case where the value is too low, and in contrast, many objects may be missed when the value is too high.  

At the $n^{\text{th}}$ iteration of the FISTA algorithm, the residual image $R^{n}$ is calculated by 
\begin{equation}
R^{(n)} = \mu \mathbf{A}^{\text{T}} (V - \mathbf{A \Phi}\alpha^{(n)} )
\label{FISTAresidual}
,\end{equation}
where $V$ are the measured visibilities, $\alpha^{(n)}$ are the coefficients in the dictionary $\mathbf{\Phi}$ of the solution at the $n$$^{\text{th}}$ iteration, and $\mu$ is the FISTA relaxation parameter that depends on the matrix $ \mathbf{F=A \Phi}$ ($\mu$ must verify $0 < \mu <  \frac{1}{\parallel F \parallel}$).
If the chosen dictionary $\Phi$ is the starlet or the curvelet transform, then the noise level can be estimated for each scale $j$ of the residual $R^{(n)}$ using a reliable estimator such as the MAD (median of the absolute deviation):
\begin{equation}
\sigma_j^{(n)} = \frac{MAD  \left(  \alpha_j^{(R^{(n)})}  \right) }{0.6745}
\label{MAD}
,\end{equation}
denoting $\alpha_j^{(R^{(n)})}$ the coefficients of $R^{(n)}$ of the band $j$ in the dictionary $\mathbf{\Phi}$. The thresholds $\lambda_j$ (see Eq. \ref{minimize_l0_pix_2}) are then different for each band, with $\lambda_j = k \sigma_j$, where $k$ fixes the probabilities of false detections. This provides the advantage of properly handling correlated noise. The constant factor in Eq. \ref{MAD} is derived from the quantile function of a normal distribution taken at probability 3/4 \citep{madref}.

 \begin{algorithm}[!htb]
\caption{FISTA implementation}
\label{Fista}
\begin{algorithmic}[1]
\REQUIRE $\quad$ \\
A dictionary $\bm{\Phi}$.\\
The implicit operator $\mathbf{A}$.\\
The original visibility data $V$.\\
A detection level k.\\ 
The soft-thresholding operator:\\
$\text{SoftThresh}_{\lambda}(\mathbf{\alpha})=\prox_{\bm{\lambda}}(\mathbf{\alpha})=\left(\left(1-\frac{\lambda_j}{|\alpha_j|}\right)_{+} \alpha_j \right)_{1\leq j\leq N}$
        
\bigskip

\STATE {Initialize $\alpha^{(0)}$=0,  $t^{(0)}$=0}

\FOR{$n=0$ to $N_{\max}-1$} 
 \STATE {$\bm{\beta}^{(n+1)} = \mathbf{x}^{(n)} + \mu \bm{\Phi}^{\text{T}} \mathbf{A}^{\text{T}}(V-\mathbf{A}\bm{\Phi}\bm{\alpha}^{(n)})$}
 
 \STATE{$\mathbf{x}^{(n+1)}=\text{SoftThresh}_{\mu \bm{\lambda}_{j}} \bm{\beta}^{(n+1)}$}
 
 
  \STATE {$t^{(n+1)}=(1+\sqrt{1+4 (t^{(n)})^2})/2$}
    \STATE {$\gamma^{(n)}=(t^{(n)}-1)/t^{(n+1)}$}
        \STATE {$\bm{\alpha}^{(n+1)}=\mathbf{x}^{(n)}+\gamma^{(n)}(\mathbf{x}^{(n)}-\mathbf{x}^{(n-1)})$}
 \ENDFOR
 
 \STATE {\bf Return:} image $\bm{\Phi\bm{\alpha}}^{(N_{\max})}$.
\end{algorithmic}
\end{algorithm}

\section{Benchmarking of the sparse recovery}
\label{benchmark}
\subsection{Benchmark data preparation}
\label{Preliminaries}

The performance of CS can be determined and compared to that of other imaging methods such as CoSch-CLEAN and MS-CLEAN. We used here three different LOFAR datasets (Table \ref{tabdatasets}) to compare CS and CLEAN-based algorithms in simulated and real situations.

\begin{table*}
\begin{center}
\begin{tabular}{ |c|c|c|c|}
\hline
        \textbf{Name} & \textbf{A} & \textbf{B} & \textbf{C}\\ \hline
        Sources & Two point sources & Source grid \& W50 & Cygnus A \\ 
        
        Nature of the data & simulated & simulated & real \\ 
        Pointing RA (hms) & 02$\text{h}$45$\text{m}$00.0''  &14$\text{h}$11$\text{m}$20.9$\text{s}$  &19$\text{h}$59$\text{m}$28.0$\text{s}$ \\ 
        Pointing DEC (dms) & +52$^\circ$54$\text{m}$54.4$\text{s}$ & +52$^\circ$13$\text{m}$55.2$\text{s}$ & +40$^\circ$44$\text{m}$02.0$\text{s}$ \\ \hline

        Station configuration & core& \textit{Core+Remote} & \textit{Core+Remote} \\ 
        $N_{\text{stations}}$ & 48 & 55 & 36 \\
        $N_{\text{correlations}}$ & 1176 & 1540 & 666 \\ 
        $UV_{max}$ ($k\lambda$) & 3.7 $k\lambda$ & $\sim$65 $k\lambda$ & $\sim$39 $k\lambda$ \\ 
        Maximum resolution (') & $\sim$1$'$  & $\sim$3$''$  & $\sim$5$''$  \\ \hline
        Polarization & \multicolumn{3}{c|}{(linear) XX, YY, XY, YX} \\ \hline

        $t_{start}$ (UT) & 6-Nov-2013/00:00:00.5 & 13-Apr-2012/18:25:05.0 &2-Mar-2013/05:50:50.0 \\ 
        $t_{end}$ (UT) & 6-Nov-2013/00:59:59.5 & 14-Apr-2012/04:44:56.6 &2-Mar-2013/11:50:40.0 \\ 
        $\delta t$ (s) & 1 & 10.01 & 10.01 \\ 
        $N_{\text{times}}$ & 3600 (=1 Hr) &37192 ($\sim$10 Hrs) &21590 ($\sim$6 Hrs) \\ 
        $\nu_0$ (MHz) & 150.1 & 116.2 & 151.2 \\ 
        $\Delta\nu$ (MHz) & 0.195 & 0.183 & 0.195 \\ \hline
Section & \S\ref{angularsep} & \S\ref{WFsourcegrid} \& \S\ref{secw50} & \S\ref{seccygA} \\ \hline
\end{tabular}
\caption{Summary information of the LOFAR datasets used in the present study. $N_{\text{stations}}$ and $N_{\text{correlations}}$ are the number of stations and the corresponding number of independent correlation measurements (computed as $N_{\text{correlations}}=N_{\text{stations}}(N_{\text{stations}}-1)/2 + N_{\text{stations}}$). Autocorrelations were systematically flagged before imaging. $UV_{max}$ is the largest radial distance in the (u,v) plane that limits the maximum theoretical resolution of the instrument. $t_{start}$ and $t_{end}$ observation times are in Universal Time (UT), providing a total number of time samples $N_{\text{times}}$ at a time resolution $\delta t$. We considered only one frequency channel covering a bandwidth $\Delta \nu$ centred on the frequency $\nu_0$. All three datasets come in \textit{measurements sets} that follow the NRAO standards. Dataset \textbf{A}  has been generated for a one-hour synthetic zenith observation using \textit{mkant} and \textit{makeMS} routines that are part of the LOFAR software package. Dataset \textbf{B} comes from a LOFAR commissioning observation and is filled with simulated data. Dataset \textbf{C} contains actual calibrated data from a real LOFAR observation of Cygnus A.}
\label{tabdatasets}
\end{center}

\end{table*}




We implemented CS by creating a new method in \textit{AWimager} based on the current CoSch-CLEAN implementation. This new method serves as an entry point to the LOFAR imager to connect an external library (developed by the authors, see Sect. \ref{soft}) containing the dictionaries (wavelets and curvelets), transforms, and minimization methods described above.
The CS cycle involves a continual passing of data back and forth between \textit{AWimager} and the library (before and after \textit{gridding}, \textit{degridding} of the data, and applying the \textit{A-} and \textit{W-projections}). At step \textit{i} in the CS cycle, \textit{AWimager} performs the following operations:
 \begin{itemize}
 \item it takes as input an image reconstruction obtained at iteration $i-1$ (represented in the selected dictionary),
 \item it simulates the visibilities using the same $\mathbf{A}$ operator as for CoSch-CLEAN (and perform \textit{degridding}),
 \item it compares the simulated visibilities to the original ones, and
 \item it passes the difference back to the dictionary using $\mathbf{A^T}$, and performs a step of the CS minimization, thresholding, and produces a reconstructed image of iteration $i$.
 \end{itemize}
 
The process stops if the maximum number of iterations is reached or if a convergence criterion (based on the noise of the residual) is met. 
The current implementation fits the current framework used by recent algorithms where the major cycle performs the CS operations (especially line 3 of Algo \ref{Fista}, which enables computing Eq. \ref{FISTAresidual}). In the future, multiple steps of CS will be possible (in a minor cycle) to improve performance.
\textit{AWimager} outputs a similar set of files for both CoSch-CLEAN and CS (the restored, residual, model, and a PSF image). CLEAN output is a CLEAN-beam-convolved image; in the case of CS, the reconstructed image is the solution $\mathbf{\Phi} \alpha^n$ (Eq. \ref{FISTAresidual}). 
The program user may choose to convolve the compressed sensing output with the CLEAN beam and/or add the residual image (see discussion in Sect. \ref{convolution}). 
The CS residual is $R^{(n)}$ (Eq. \ref{FISTAresidual}) at the last iteration of CS and is analogous to the residual at the output of x-CLEAN.
%
By selecting the CS method, the meaning of some classical imager parameters are changed in the scope of the previous definitions: the gain in x-CLEAN controls the fraction of the PSF subtracted at each iteration. In CS, the gain is the relaxation parameter $\mu$ of Eq. \ref{FISTAresidual}, which controls the convergence rate of the algorithm. The threshold value in x-CLEAN is a flux density value usually associated with $n$ times the level of noise measured (or expected) in the residuals (or in an source-free patch of the dirty image). Setting this threshold to zero would basically lead x-CLEAN to false detections from the background features. 
As discussed above in Sect. \ref{AlgoMAD}, the CS thresholding parameter is defined for each or all bands in the chosen dictionary.
The number of iterations for CoSch-CLEAN algorithm is set to a high value ($>10^4$ to set an unrestrained convergence to the threshold). The number of iterations for CS can be set to an arbitrary value or be controlled by a convergence criterion.

The following figures of merit will measure the quality of the reconstruction using CS and CoSch-CLEAN methods:
\begin{itemize}
\item the total flux density computed in a region \textbf{S} around the source,
\item the residual image standard deviation (Std-dev) and root mean square (r.m.s.) computed over {\textbf S} in the residual image or over the full image \textbf{I},
\item the error image and the mean square error (MSE) computed over \textbf{I} when a model image is available,
and 
\item the dynamic range (DR) defined as the ratio of the maximum of the image to the residual Std-dev.
\end{itemize}

We here present different benchmark reconstructions arranged in increasing source complexity.
In Sect. \ref{unresolved}, we reconstruct point sources in two typical situations implied by new-generation radio interferometers: at small scale, by reconstructing the image of two point sources at different angular separation (Sect. \ref{angularsep}),
and at large scale, by reconstructing high-dynamic and wide-field images using a grid of point sources (Sect. \ref{WFsourcegrid}). In Sect. \ref{resolved} we monitor the reconstruction quality of extended emissions: from a simulated W50 observation (Sect. \ref{secw50}) and a real LOFAR observation of Cygnus A (Sect.\ref{seccygA}).

\subsection{Unresolved sources}
\label{unresolved}
\subsubsection{Angular separation between two sources}
\label{angularsep}
When considering point sources, the x-CLEAN algorithms are usually the most frequently adapted reconstruction method. The first test is therefore to compare the performance of CS reconstruction to that of CoSch-CLEAN for that simple type of source. 
We generated empty datasets (dataset \textbf{A} in Table \ref{tabdatasets}) describing a simulated one-hour LOFAR observation. We only used core LOFAR stations, since they are included in most LOFAR observation. This justifies the evaluation of the ground performance of CS on this array subset. The layout of the core stations is slightly elongated ($\sim$10\%, \cite{2013arXiv1305.3550V}) in the north-south direction, providing a symmetric beam pattern to a direction close to zenith. In this dataset, we pointed the array to the local zenith, which will induce an a priori elongated beam pattern.
The core stations in dataset \textbf{A} give a highest resolution of 1$'$ at $\nu_0$=150 MHz (in the HBA band).In addition, we restricted the (u,v) coverage to the radial distances $[r_{(u,v)}^{min}$,$r_{(u,v)}^{max}]$= $[$0.1 k$\lambda$, 1.6 k$\lambda]$ to artificially impose an angular resolution of $\sim$2$'$ (which is effectively close to $\sim$2.8$'$ as deduced from the PSF).

We simulated two one-Jansky point sources at frequency $\nu_0$. With this simple sky model, we used the \textit{blackboard self-cal} program (\textit{BBS}, \citet{BBS}) to fill the LOFAR dataset with simulated visibilities.
One source is located at the phase centre, the second is located at varying angular distances $\delta\theta$ ranging from 10$''$ up to 5$'$. As a consequence, we span various angular separation across the instrumental limit given by the PSF. We can define three different regimes where the performances of CLEAN and CS are compared: i) sources are unresolved, ii) partially resolved, or iii) unambiguously resolved. As we are close to the phase centre, the effects mentioned in Sect. \ref{DDE} are negligible (which will no longer be the case in Sect. \ref{WFsourcegrid}).
For different angular separations, we injected Gaussian noise on the visibilities to obtain effective S/N levels of $\sim$3, $\sim$9, $\sim$16, and $\sim$2000 (noise-free case) in the image. 

We used the original CoSch-CLEAN and our CS method implemented in \textit{AWimager}, which serves here as a common testing environment for this study. We performed $10^4$ iterations for the CoSch-CLEAN and 200 iterations for CS. As different levels of noise were simulated, we set the $k$ parameter (Eq. \ref{MAD}) to three times the noise level. We used for both algorithms the (u,v)-truncated dataset as described above and produced 512$\times$512 pixel images at 5$''$ pixel resolution. We imposed a \textit{natural} weighting scheme to improve the S/N of the resulting images.

\begin{figure*}
\centering
\includegraphics[scale=0.7]{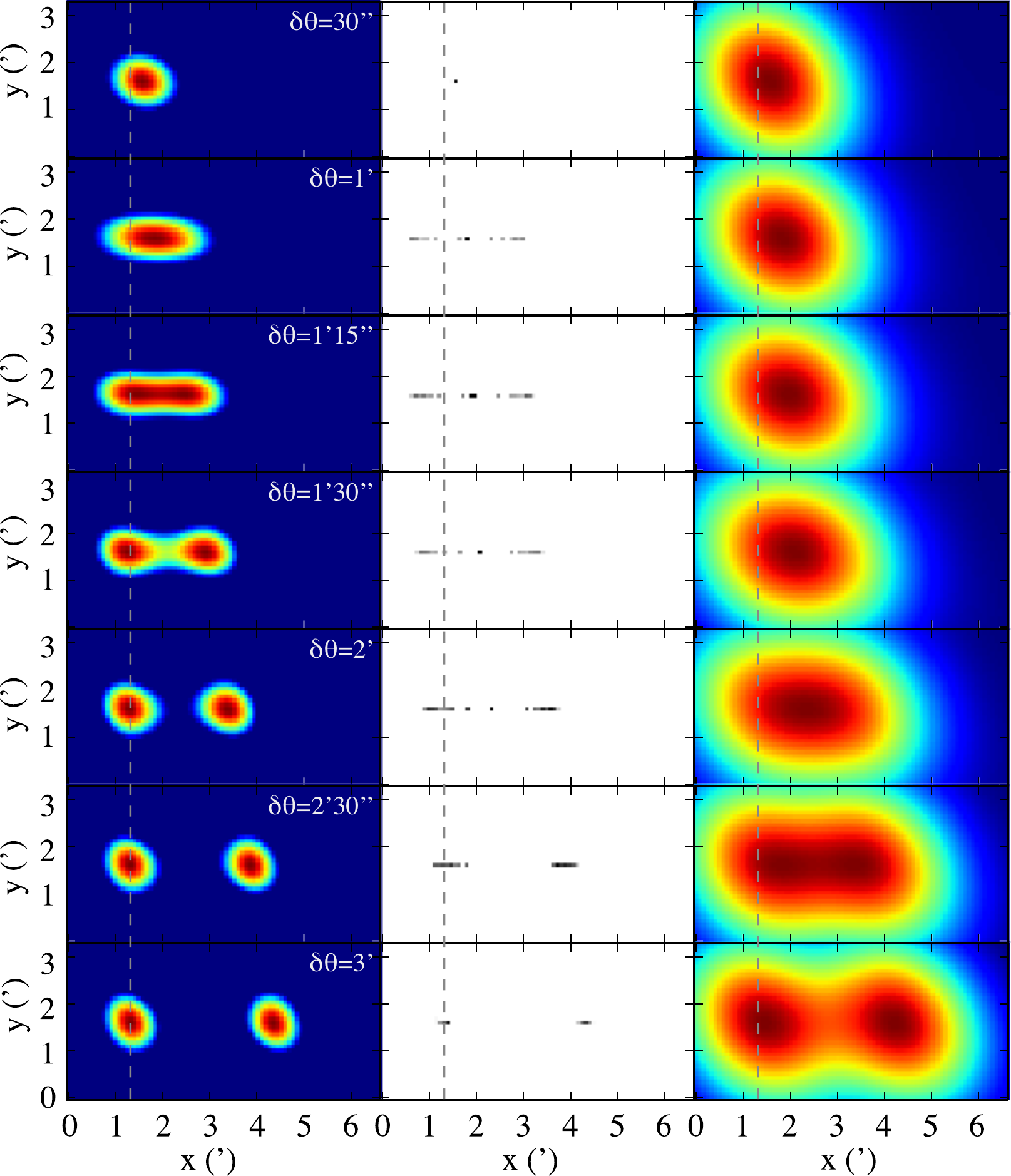}
\caption{Resulting reconstructed CS image (left column), CoSch-CLEAN model (middle column), and convolved (right column) images obtained from a simulation of two point sources with different angular separation $\delta\theta$ from 30$''$ to 3$'$ by steps of 30$''$ (from top to bottom) plus the $\delta\theta$=1$'$15$''$ case. The third column was obtained by convolving the CLEAN components (middle column) with the CLEAN beam of FWHM 3.18$'$$\times$2.55$'$. The effective separation of the two sources, after imaging by the different methods, occurs at smaller angular separation with CS (between $\delta\theta$=1$'$ and $\delta\theta$=1$'$15$''$) than with CoSch-CLEAN (between $\delta\theta$=2$'$ and $\delta\theta$=2$'$30$''$). The unambiguous separation of the two sources was obtained within few pixel errors, starting from $\delta\theta$=1$'$30$''$ for CS and $\delta\theta$=3$'$ for CoSch-CLEAN. Each cropped image was originally of size 512$\times$512 pixels of 5$''$. The (u,v) coverage has been restricted to the [0.1 k$\lambda$,1.6 k$\lambda$] domain to enforce an artificial resolution of $\sim$2$'$. The colour map scales are normalized to the maximum of each image, but colour bars were not represented for clarity. The contrast of the CLEAN components (middle column) was enhanced to ease their visibility. The grey vertical dash line marks the true position of the source located at the phase centre of the dataset.}
\label{gridangsep}
\end{figure*} 

Figure \ref{gridangsep} illustrates the results obtained with the noise-free dataset: the CS reconstruction (left column), the CoSch-CLEAN model image (middle column), and the corresponding \textit{convolved} image (right column). The rows correspond to seven values of angular separation $\delta\theta$ from 30$''$ to 3$'$ by steps of 30$''$ and the $\delta\theta$=1$'$15$''$ case.
For the CS reconstruction, there is no model image in the CLEAN sense because the output of CS is directly the best representation of the true sky at a finite resolution (similar to the comparison of models in \citet{Li2011}).
Different angular separation criteria can be chosen. One can apply the Rayleigh criterion based on the separation of two (pixel) CLEAN components on the model image or use source-finders on reconstructed images (see below).
Here (and hereafter), the CLEAN beam, setting the highest angular resolution of CLEAN deconvolved images, has a size of 3.18$'$$\times$2.55$'$ (major$\times$minor axes).
In the CoSch-CLEAN model image, when $\delta\theta$ is small, only one group of CLEAN components is detected at the mean position of the two sources. Starting from $\delta\theta$=1$'$30$''$, additional CLEAN components are being detected on the \textit{wings} of the main component, but are located correctly on the source position. With $\delta\theta$ increasing, the amplitudes of these side CLEAN components increase and contribute in the resulting elongated shape on the convolved CLEAN image. Starting from $\delta\theta$=2$'$30$''$, the two sources are unambiguously separated into two distinct groups of components around the correct position. The position uncertainty of these two sources is still high ($\sim$30$''$). For separations larger than $\delta\theta$=3$'$, the astrometric and flux density errors start to decrease. In the regime where CoSch-CLEAN cannot unambiguously resolve the two sources, the features present in the model images (the central components as well as the wing components) cannot be associated with real sources. In the scope of this method, an exploitable image with limited resolution can only be obtained after convolving with the CLEAN beam.

With CS, we directly obtain the best estimate of the sky, as shown in the reconstructed image in the left column of Fig. \ref{gridangsep}. In the partially resolved regime, we also note that the elongation of the source size occurs at $\delta\theta$=1$'$. We note that the two point sources are resolved at lower angular separation ($\delta\theta$$\sim$1$'$15$''$) than for CoSch-CLEAN. An elliptical fit of the FWHM of the sources gives the source size, which can be seen as the \textit{effective} CS convolution beam. In this case, its dimensions are 1.55$'$$\times$1.09$'$, representing a beam of cross-section approximately 1.39$'$, which is smaller than that of the CLEAN beam.
In addition to the improved angular separability, the CS reconstructed sources are also correctly located in the image (e.g. the central source lies close to the vertical mark in Fig. \ref{gridangsep} for $\delta\theta\leq$1$'$30$''$) as compared to the CoSch-CLEAN model image, where the CLEAN components are only correctly positioned starting from $\delta\theta\leq$2$'$30$''$. 
In the noise-free case, by using exactly the same dataset and imaging parameters, these results suggest that CS is able to recover information on the true sky beyond the theoretical resolution limit (which is constant in this dataset).
Because there is always a non-zero level of noise in the calibrated interferometric data, we tested the reliability of this characteristic against the image S/N by using the different noisy datasets mentioned above. 

To compute the effective angular separability of the sources using CoSch-CLEAN and CS, we used the LOFAR pyBDSM\footnote{See http://www.lofar.org/wiki/doku.php for more information.} package (\textit{python blob detection and source measurement}), which consists of island detection, fitting, and characterization of all structures in the image. We defined a detection threshold so that no other artefact could be detected as a source in the images because we only focused on the two simulated point sources. We used the same set of detection parameters for CoSch-CLEAN and CS images.

At a specific S/N level, the limit value of angular separation at which two distinct sources can be distinguished by the source finders gives an estimate of the separability angle of the sources and therefore of the effective resolution that limits the typical size of genuine physical features. The CoSch-CLEAN beam and the effective CS beam are not circular, therefore the absolute effective angular resolution may depend on the orientation of these beams with respect to the direction joining the two sources. To have a measure that is independent of this direction, it is probably better to compute the ratio between the beam elliptical parameters obtained with CoSch-CLEAN and that obtained with CS.

\begin{center}
\begin{figure}[!h]
\centering
\includegraphics[scale=0.55]{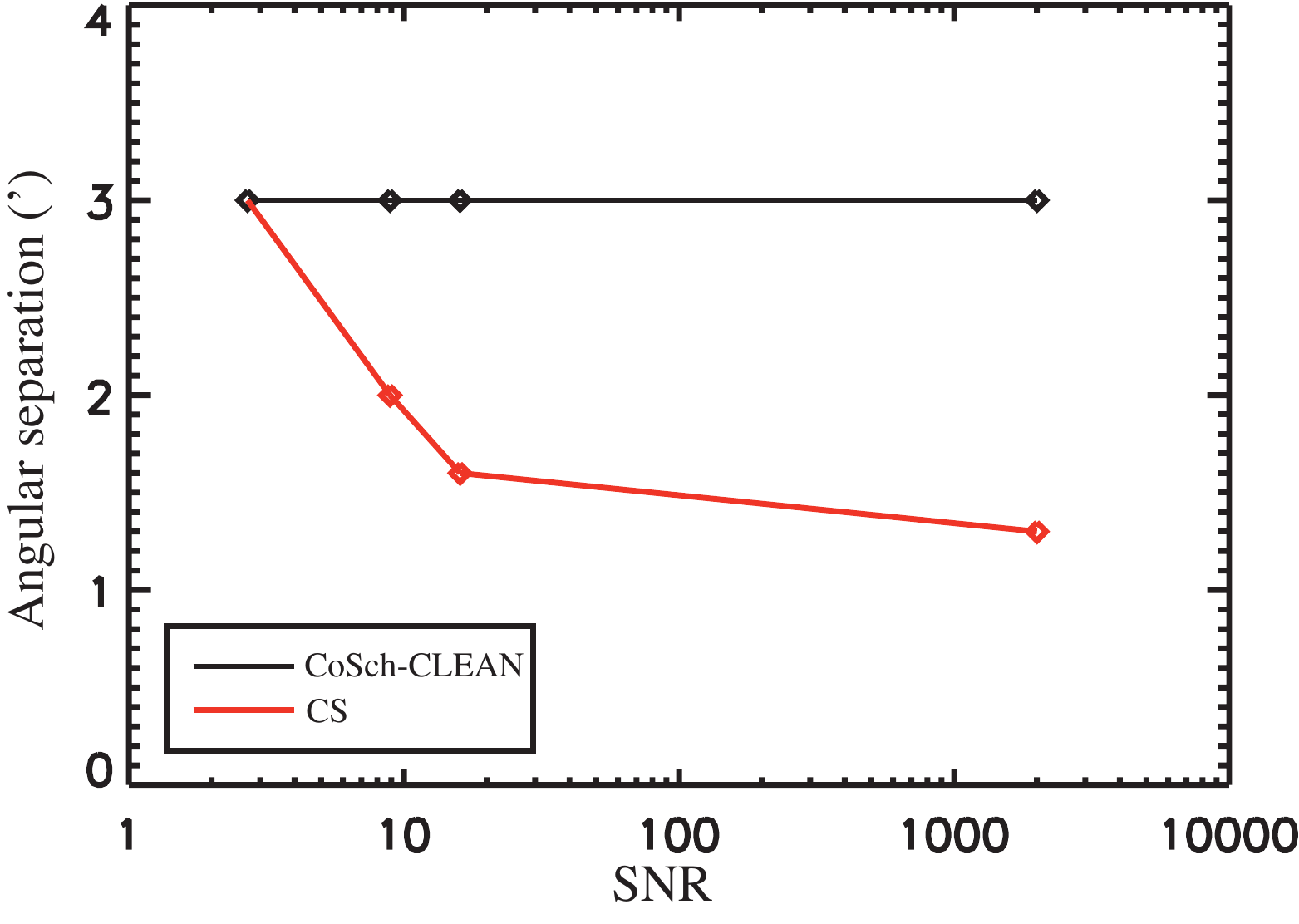}
\caption{Values of $\delta\theta_{min}$ of source separability as a function of the S/N for CS (red curve) and CoSch-CLEAN (black curve). In the moderate (S/N $\geq$5) and high S/N regimes, the resulting source separability is improved by a factor 1.3 to $\lesssim$2.}
\label{plotangsep}
\end{figure} 
\end{center}

Using an appropriate sampling of $\delta\theta$ and noise levels, we can build a graph of this effective resolution deduced from the separability of the two sources for different S/N levels. We present in Fig. \ref{plotangsep} the resulting curves obtained with CoSch-CLEAN (black) and CS (red). For each S/N we derived the effective improvement brought by CS by noting the angle $\delta\theta_{min}$ at which the two sources are separated.
On the one hand, the CoSch-CLEAN separability has a limited dependence on the S/N, which makes it a stable and reliable algorithm for detecting point sources in various S/N regimes. On other hand, CS behaves differently. At a high S/N regime, the separability values outperforms that of CoSch-CLEAN by a factor of 2--3 in the [10,2000] S/N range. When the S/N decreases, the separability of CS tends to that of CoSch-CLEAN.


In terms of astrometric error, the detected point source locations can be compared to the ($\alpha$,$\delta$) coordinates of the input sky model (where the sources were placed at pixel centres). The relative position errors do not exceed 1$'$ for CS and 3$'$ for CLEAN for all the different noise levels.

In addition to the angular separability and astrometric errors, we inspected the flux density of the source. 
Naturally, the flux density error of the reconstructed sources is affected by the level of the background noise and scales with the S/N. It is found to be 3\% in the low-noise regime and up to 25\% in the high-noise regime for CS and 3\% to 23\% for CoSch-CLEAN.

From this perspective, it appears that CS provides results that are almost as good as the results by CLEAN by default, and provides an improved angular separability with high and moderate S/N data. 
The low astrometric and flux density error confirm the superior resolution capability of CS, which suggests the possibility of dramatically improved angular resolution of extended emission from poorly sampled interferometric data.


\subsubsection{Wide-field imaging of a grid of sources}
\label{WFsourcegrid}

\begin{figure}
\begin{minipage}[!t]{.48\textwidth}
\centering
\includegraphics[scale=0.40]{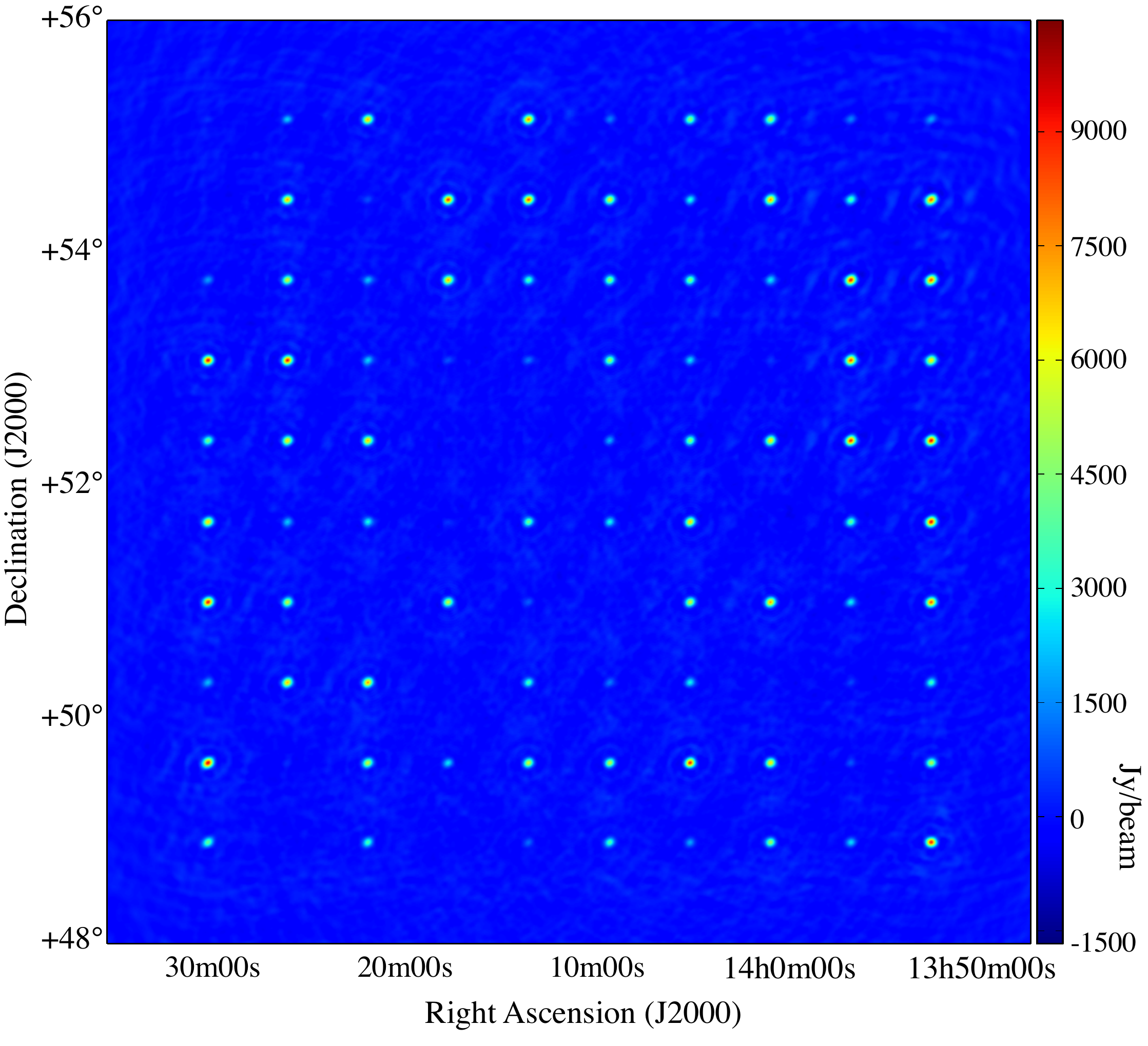}
\caption{Dirty image derived from a set of 100 simulated point sources with flux density ranging from 1 to $10^4$ Jy. Simulated visibilities are generated with dataset \textbf{B}. Some sources are not yet visible because of the dynamic of the source and the noise level.}
\label{point_sources}
\end{minipage}
\end{figure}

Interferometry imaging at low frequencies implies a larger FoV because of the size of the LOFAR station analogue beam. By using \textit{AWimager}, which deals with wide FoV \citep{tasse13}, we checked the ability of CoSch-CLEAN and CS to recover the correct flux density of point sources that are away from the phase centre.
We simulated a 10$\times$10 grid of point sources over a region of 8$^{\circ}\times$8$^{\circ}$ around the phase centre. Their flux densities range from 1 to $10^4$ Jy. We used BBS to fill dataset {\textbf B} by enabling the simulation of the beam (for the \textit{A-projection}). The \textit{W-projection} only depends on the layout of the interferometer, which is included in the dataset. Noise was injected into the dataset to provide an r.m.s. value of $10^{-4}$ (so $\sim$1 Jy) relative to the peak of the dirty image. The points are located at equidistant vertices in the grid, which enabled us to examine the distribution of the flux density and the potential residual distortions over the field without overlapping of the source. While this arrangement is unrealistic, we can still monitor the astrometric and flux density accuracy versus the radial distance from the phase centre. We applied CoSch-CLEAN and CS to the simulated dataset containing the grid. The dirty image generated from visibilities is shown in Fig. \ref{point_sources}.

\begin{figure}
\centering
\begin{minipage}[!t]{.48\textwidth}
\includegraphics[scale=0.48]{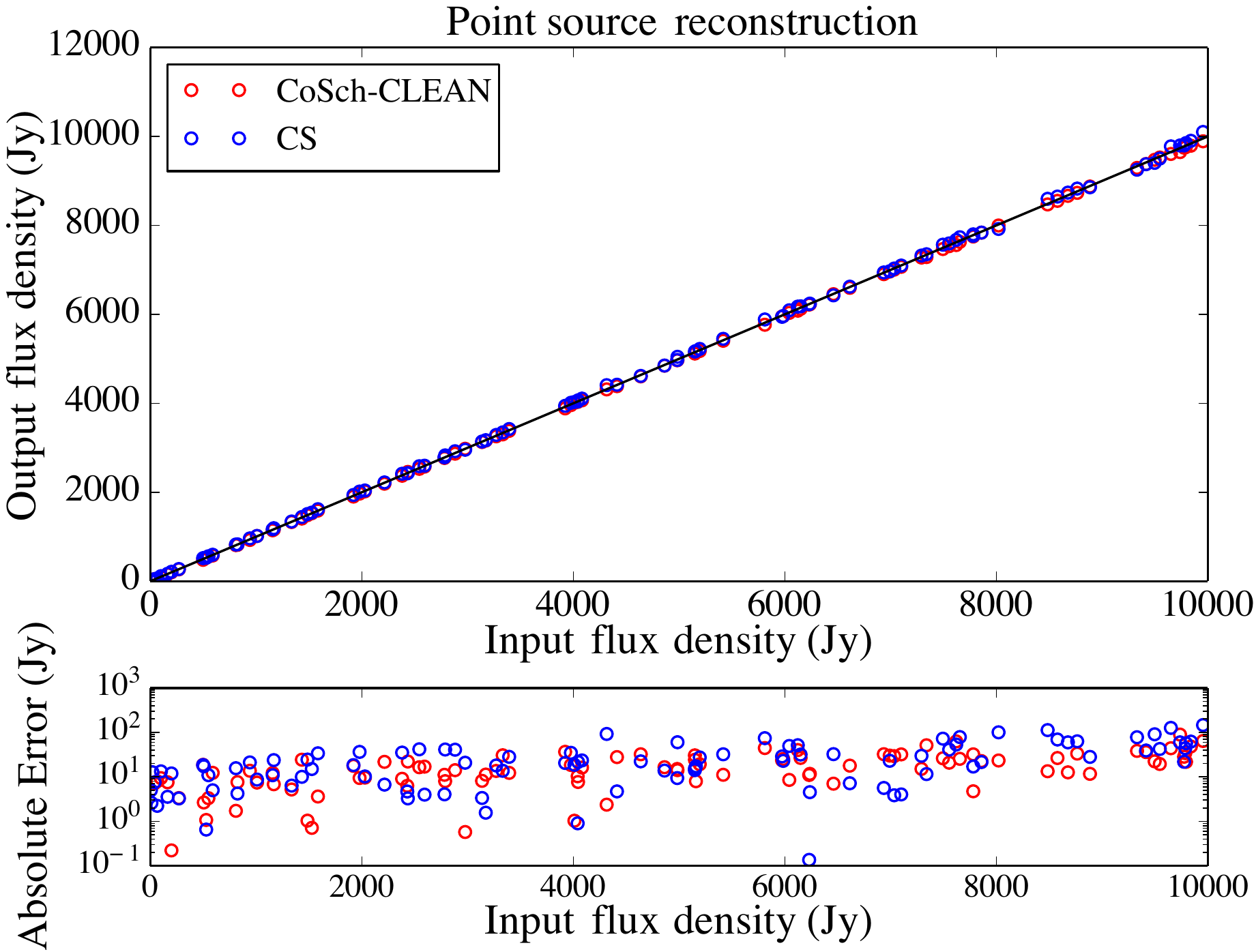} 
\caption{(Top) Total flux density reconstruction for a set of point sources (Fig. \ref{point_sources}) with original flux density spanning over 4 orders of magnitude with the original flux density (x-axis) vs. the recovered flux density (y-axis). (Bottom) scatter plot of the absolute the error for each source. The recovered flux densities for CoSch-CLEAN (red) and CS (blue) are represented on a linear scale, whereas the absolute error is on a logarithmic scale for clarity. Perfect reconstruction lies along the black line. The output flux density values and errors are similar with both CoSch-CLEAN and CS.}
\label{points_results}
\end{minipage}
\end{figure}

We used 1024$\times$1024 pixel images with a pixel size of 28$''$ and the full (u,v) coverage of dataset \textbf{B} was used for imaging, giving an effective angular resolution of $\sim$3$''$.
For CoSch-CLEAN, we used $10^8$ iterations and for CS, we used 200 iterations. With CS, the sources were reconstructed with the starlet dictionary.
We report in Fig. \ref{points_results} the output integrated flux density of all detected sources in the reconstructed images against the input flux densities of the sky model point sources. A sound reconstruction should put every source on the first bisector. We again used pyBDSM to perform the source detection and characterization (including the 2D elliptical Gaussian fit of the source, position of the Gaussian barycentre, and photometry along with providing respective errors). The error bars of each point were obtained directly from the photometry and are negligible given the low level of the background noise. However, these error bars do not include the bias of poorly reconstructed sources.

For the CS reconstructed image, the effective source size is reduced to smaller patches with a few pixels each around the sources (comparable to the CLEAN model, only constituting a collection of pixels). The flux is more efficiently gathered around the source position, resulting in an increase of all pixel size values in the CS images (in Jy/beam). 
The resulting source size was 3.2$'$$\times$3.7$'$ (i.e. the dimension of the CLEAN beam) for CoSch-CLEAN and between $\sim$30$''$--1$'$ (1-2 pixels) for CS. This result corroborates that of the previous study. The superior resolution here yields an improvement of
a factor of 3 to 6  of the source size compared to the size of the CLEAN beam, using exactly the same dataset (and weighting) for the two methods.

The flux density r.m.s. error was derived from the residual images and accounts for 3.6 Jy/beam Std-dev of CS and 1.7 Jy/beam Std-dev for CoSch-CLEAN. The error bars are not reported in the plot for clarity. The relative error (compared to the input flux density of each source) is not larger than 10\% in most cases for both CoSch-CLEAN and CS. The flux density error slightly increases with the source flux density, as depicted by the scatter plot represented in a log-scale in Fig. \ref{points_results} (bottom).

As a result of the many strong sources, CoSch-CLEAN and CS were unable to reconstruct some faintest sources barely above the background level. CoSch-CLEAN presents slightly better performances for reconstructing correct flux densities with a lower error (the mean absolute error is 19 Jy for CLEAN and 29 Jy for CS). Nevertheless, CS led to the detection of more faint sources that were missed by CoSch-CLEAN, but with a larger error on their flux density values. In spite of its improved angular resolution and its detection capability, CS provides a larger Std-dev (3.6 Jy/beam) on the residual image than CoSch-CLEAN (1.7 Jy/beam). This could be an effect of the thresholding taking place in CS or the choice of the dictionary, which is not perfectly fitted to represent point sources.




We did not note any particular dependence on the radial distance from the phase centre for either CoSch-CLEAN or CS. This suggest that the \textit{A-} and \textit{W-projections} in \textit{AWimager} correctly prevented a radial dependence of the flux, the astrometric error, and the source distortion (the last two being at the scale of one pixel). This tell us that the CS method is effectively compatible with the RIME framework. Moreover, x-CLEAN algorithms remain competitive with CS because the spatial extension of wavelet atoms is not as efficient as Dirac atoms (pixel) in representing single point sources.
%


\subsection{Extended sources}
\label{resolved}
We have shown that CS and CLEAN are clearly competing when imaging point sources. We now address the reconstruction of extended radio emission, which are not sparsely represented in a pixel dictionary. We therefore continue to use the starlet dictionary for the CS reconstruction to ensure the best sparsity of the signal. We first study a simulation of W50 (Sect.\ref{secw50}) and then discuss the results of CS on a real LOFAR observation of Cygnus A (Sect.\ref{seccygA}).

\subsubsection{Simulated observation: the W50 nebula}
\label{secw50}
\begin{figure*}
\centering
\includegraphics[scale=0.35]{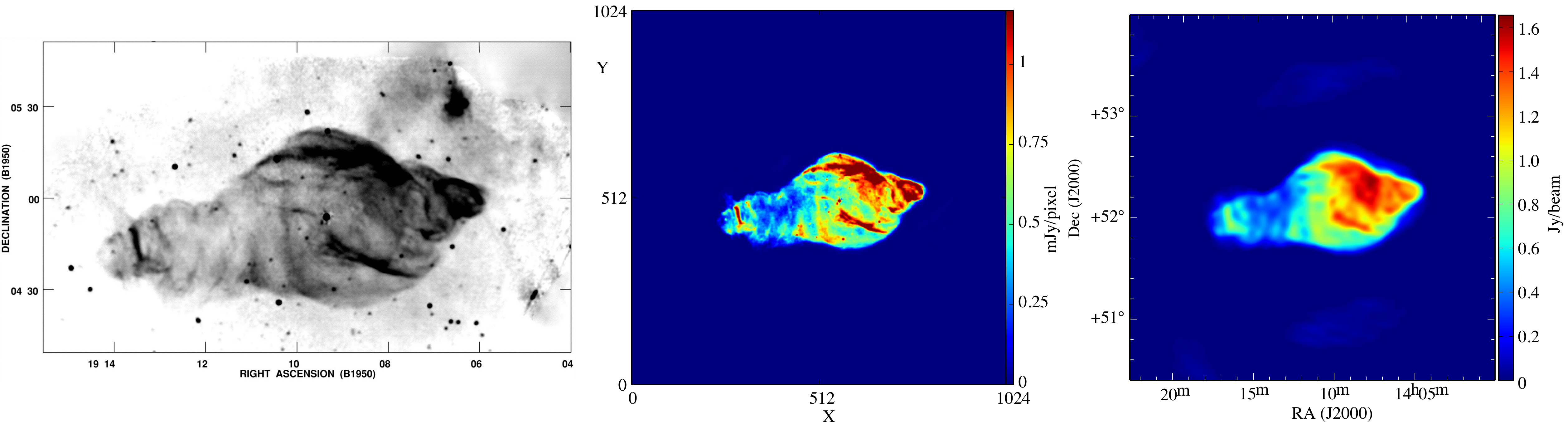}
\caption{(left) Original 1.4 GHz radio image from \citet{dubner98} (pixel size = 6.75$''$, ang. resolution = 55$''$), (middle) prepared input model image scaled to the approximate total flux density of 230 Jy (1024$\times$1024 with pixel size of 13.5$''$), and (right) the dirty image obtained with dataset \textbf{B}. All emission falling outside the extended emission of W50 was set to zero, but point sources inside W50 were kept in the simulation.}
\label{w50intro}
\end{figure*}

The W50 nebula (hosting the SS 433 microquasar) is an extended supernova remnant of large dimension ($\sim$2$^\circ\times\sim$1$^\circ$) with internal filamentous structuring that makes it proper for benchmarking CS and CLEAN-based algorithms. First, we used the W50 brightness image from \citet{dubner98} (Fig. \ref{w50intro} left). This image at 1.4 GHz is rich in information at various angular scales down to an angular resolution of $\sim$55$''$ (the image at 327.5 MHz, taken with the VLA in the D configuration, was also available, but no structures smaller than 70$'$ are resolved). We assumed that the higher spatial frequency features also emit in the LOFAR band (see W50 observed with LOFAR HBA in \citet{broderick2014} and previous work). We focused on the central extended emission by removing part of the extraneous foreground and background features. We set the flux scale of the total model image to match a total flux density of $\sim$250 Jy at 116 MHz as interpolated from \cite{dubner98}. Second, we used the \textit{predict} task of \textit{AWimager} to convert the input model to visibilities in dataset \textbf{B}. 
No artificial DDEs were inserted in the simulation but \textit{A-projection} and \textit{W-projection} were enabled for all runs. 
Dataset \textbf{B} provides a theoretical angular resolution of $\sim$3$''$ , which is higher than the 55$''$ of the original image. We therefore expect to have good results for all three methods. The \textit{predict} step samples the image at the resolution of dataset \textbf{B}. Given the size of W50, the simulated field is 3.8$^\circ \times$3.8$^\circ$. Third, artificial Gaussian noise was added to the simulated visibilities and has an effective r.m.s. of $3\times 10^{-4}$ of the peak of the dirty \textit{noise-free} image (corresponding DR$\sim$4000).
We reconstructed images using three methods: CoSch-CLEAN, MS-CLEAN, and CS in \textit{AWimager}. MS-CLEAN was imported from the \textit{LWimager}, the standard LOFAR imager, superseded by \textit{AWimager}. 

\begin{figure*}[!t]
\centering
\includegraphics[scale=0.6]{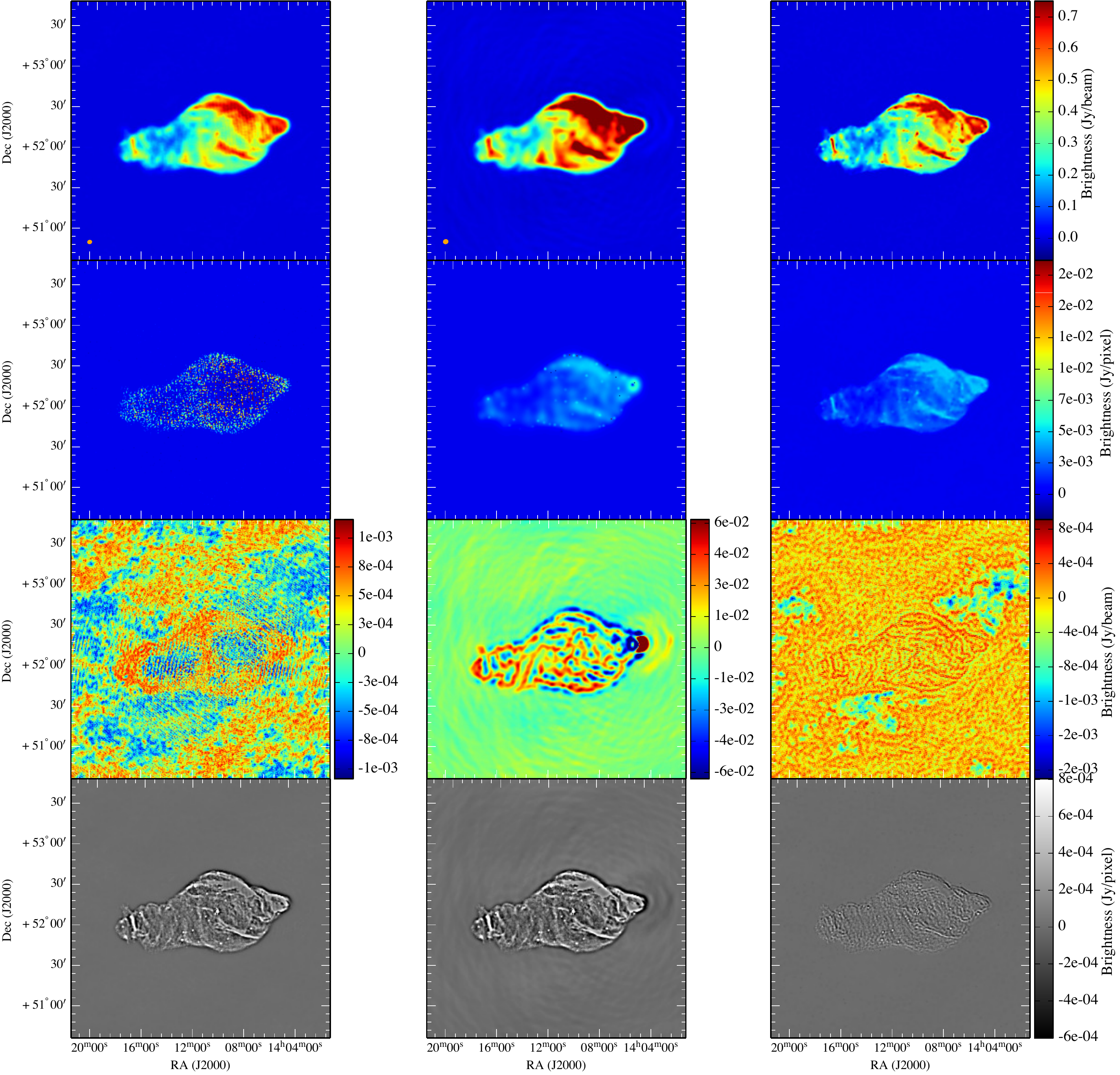}
\caption{
Reconstructed images of W50 from the simulated LOFAR observation (dataset \textbf{B}) using CoSch-CLEAN (left column), MS-CLEAN (middle column), and CS (right column). From top to bottom: the restored (in Jy/beam), the model (in Jy/pixel), the residual, and the error images. The CS restored and model images only differ by their scaling (the former is in Jy/beam using the beam area of the CoSch-CLEAN beam, the latter is in Jy/pixel). The colour scales of the images are different for each row, as indicated by the colour bar on the right. The residual images are displayed with their respective colour bar. The CS reconstruction contains high spatial frequency information restored from the dataset. The effective angular resolution of the CS image (1$'$) is close to that of the original input image (55$''$). In addition, the error image shows a closer proximity between CS and the original input image.}
\label{w50_panel}
\end{figure*}

With the input model image, we computed the error image and inspected the residual images to track the effective angular resolution and the convergence of the methods.
Table \ref{image_params} gathers all imaging parameters (as well as for Cygnus A). We compare in Fig. \ref{w50_panel} the outputs of CoSch-CLEAN (left column), MS-CLEAN (central column), and CS (right column). From top to bottom we present the reconstructed image, the model, the residual image, and the error image. The figure of merit from these images is gathered in Table \ref{summarystat}.
The total reconstructed flux densities over the source are 229.2 Jy, 229.4 Jy,  and 229.7 Jy for CoSch-CLEAN, MS-CLEAN, and CS. These values are extremely close to the 230 Jy of the model, which verifies that all three methods conserve the total flux density. This fact is a basic requirement that any new imaging method should verify.

The extended emissions were rendered with different accuracy, and a particularly good image was produced with CoSch-CLEAN because
of the low noise level in the data, the high number of iterations, and the size of the PSF. In comparison, the MS-CLEAN image presents a lower angular resolution despite taking into account the various scales of the image. Because it is a user parameter, we tried different sets of scales, thresholds, and number of iterations, but we did not improve the result, and sometimes it did not converge. This potential divergence may be caused by an inappropriate thresholding when the background residual level is reached, the algorithm then start to get signal from the background noise. In comparison, we also used the MS-CLEAN method of \textit{LWimager} and obtained equivalently poor results. MS-CLEAN results might be improvable with additional controls and masks, but we chose a straightforward imaging (as our algorithm) to highlight the robustness of our algorithm with a limited number of user controls. Moreover, the implementation of MS-CLEAN in \textit{AWimager} is still experimental and needs to be more extensively validated. This method is known to usually improve the representation of extended sources.

The restoring beam was 2.8$'$$\times$2.4$'$ for CoSch-CLEAN and 2.8$'$$\times$3.3$'$ for MS-CLEAN. The CS image is visually sharper than the other two and contains high-frequency features that were recovered during the imaging. From the typical point source size in the CS image, the effective angular resolution was $\sim$1$'$$\times$$\sim$1$'$ , nearly equal to the 55$''$ original resolution, and represents
an improvement of almost a factor 2.5-3 over x-CLEAN images.

With MS-CLEAN, the extended features are represented by large blobs in the model, along with some pixel-sized CLEAN components.
As described earlier, the output of CS comes directly as the best estimate of the sky and is in units of Jy/pixel in the model image. We multiplied the CS model image by the beam area to obtain the same brightness units (Jy/beam) as for the other images. Therefore, the model image and the restored image only differ by the residual image that was added to the latter. CS tends to concentrate even more the flux of point sources around the source (see Sect. \ref{angularsep}), and the starlet dictionary is able to correctly represent the extended emission.

Any vestigial structures in the residual image show a lower rate of convergence, which might be due to an insufficient number of iterations (which is currently not the case for the x-CLEAN algorithms), or a limitation due to the imaging method itself. In our case, the CoSch-CLEAN image, supposed to perform better with point sources than with extended emission, presents a lower residual Std-dev level ($4.1.10^{-4}$ Jy/beam) than that of MS-CLEAN ($1.3.10^{-2}$ Jy/beam) in the present situation. However, the CS residual Std-dev level is slightly better ($3.2.10^{-4}$ Jy/beam) across the image.
The error images are shown in the last row of Fig. \ref{w50_panel} and are the difference between the restored image (converted to Jy/pixel) with the initial full resolution input image (also in Jy/pixel). The input image was not degraded by convolution to match a particular resolution. From the different error images, CS visually provides the lowest error. The high-resolution features were partly restored by the CS reconstruction. The CS algorithm has the lowest mean square error value over the image ($4.8.10^{-5}$ Jy/pixel), which represents an improvement of a factor $\sim$2 compared to that of present CoSch-CLEAN and MS-CLEAN reconstructed images.

\subsubsection{Real LOFAR Observation: the Cygnus A radio galaxy}
\label{seccygA}
Cygnus A (3C 405) is one of the most powerful galaxies observable in the radio domain. It serves as a good test case for benchmarking our method: it contains extended emission, originating from two main radio lobes (representing a projected size of $\sim$2$'$$\times$1$'$) as well as compact emissions at the extremities of each lobe (the hotspots A and B in the western lobe and hotspot D in the eastern lobe, as labelled in \citet{hargrave74}). Cygnus A has recently been observed by LOFAR during its commissioning phase \citep{mckean10}, but has also been long observed in the past at low frequencies for instrumental calibration or science (e.g. \cite{laziocygA} and references therein, who observed Cygnus at 74 MHz and 327 MHz with the VLA). 
We used a real calibrated LOFAR dataset (\textbf{C}) at 151 MHz containing real noisy measurements. The data were previously calibrated on the Perley-Butler 2010 absolute flux scale (\citet{mckean10}).
\begin{table*}[!t]
\begin{center}
\vspace{3mm}
\begin{tabular}{ |c|c|c|c|c|c|c| }
\hline
 \multirow{2}{*}{\bf Algorithm} & \multicolumn{3}{c|}{{\bf W50}} & \multicolumn{3}{c|}{{\bf Cygnus A}} \\ \cline{2-7}
 & CoSch-CLEAN& MS-CLEAN  &  CS & CoSch-CLEAN& MS-CLEAN  &  CS \\ 
  \hline
  
  Specific parameter &-- &5 scales& -- & -- &5 scales & -- \\\hline  
  Weight &\multicolumn{3}{c|}{Briggs (robust = 1)} &\multicolumn{3}{c|}{Uniform} \\
  \hline 
  Planes for W-projection  &\multicolumn{6}{c|}{11} \\
  \hline  
Image size &\multicolumn{3}{c|}{1024$\times$1024} & \multicolumn{3}{c|}{512$\times$512} \\
  \hline 
  Padding & \multicolumn{3}{c|}{1.6}&\multicolumn{3}{c|}{1.5}\\
  \hline
Pixel size & \multicolumn{3}{c|}{13.5$''$}& \multicolumn{3}{c|}{1$''$}  \\
  \hline 
  Threshold &$10^{-3}$ Jy&$10^{-3}$ Jy &$10^{-3}$ & 0.5 Jy& 0.5 Jy & 0.5\\
  \hline 
  Gain & 0.01& 0.4& 0.4& 0.05 & 0.05&  0.05 \\
  \hline 
  $N_{iter}$ &$10^{8}$&  $10^{3}$ & 200 & $3.10^{4}$ & $10^{4}$& 200\\
  \hline 

\end{tabular}
\vspace{2mm}
\caption{List of relevant imaging parameters used by \textit{AWImager} for CoSch-CLEAN, MS-CLEAN, and CS, for the two extended objects. For MS-CLEAN, we have used scales at [0,10,30,70,100] pixels for W50, and [0,2,5,10,15] pixels for Cygnus A.}
 \label{image_params}
\end{center}
\end{table*}

\begin{figure*}[!ht]
\begin{center}
\includegraphics[scale=0.73]{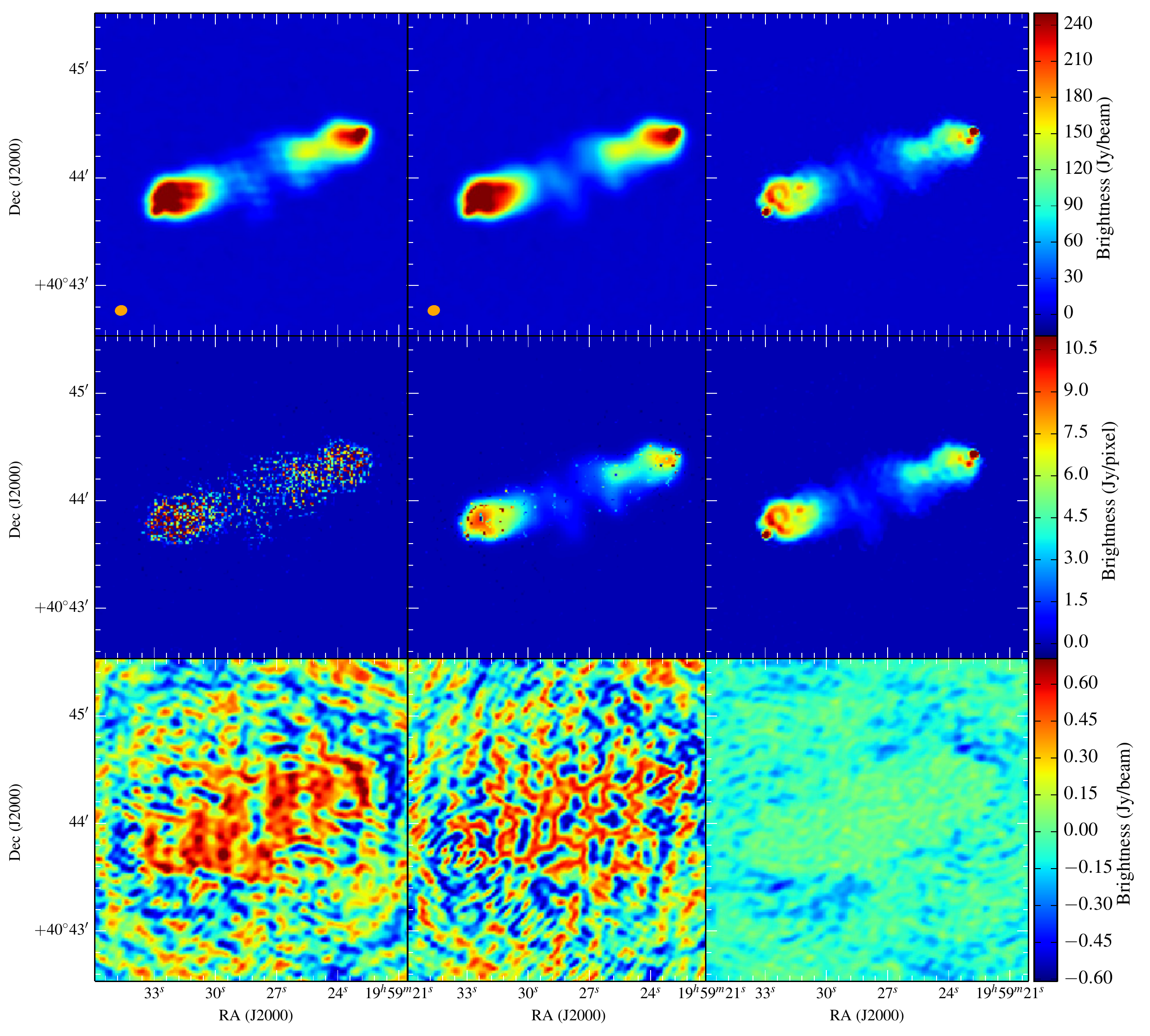}
\caption{Reconstructed images of Cygnus A from the real LOFAR observation (dataset \textbf{C}) using CoSch-CLEAN (left column), MS-CLEAN (middle column), and CS (right column). From top to bottom: the restored images, the model images, and the residual images. There is no error image, as its calculation would require a highly resolved image of the true brightness distribution of Cygnus A to compare with the recovered image. The colour scales of the images are different in each row, as indicated by the colour bar on the right. The CS reconstruction presents a higher angular resolution and a lower residual level (see text) than that obtained with the two other methods.}
\label{cygnus_panel}
\end{center}
\end{figure*}

Similar imaging parameters were used (Table \ref{image_params}). With the FoV being relatively small and the source being the dominant source in the dataset, the DDEs were expected to be low. In Fig. \ref{cygnus_panel}, we show the same output images as in Sect. \ref{secw50}. However, we do not have an input model image to compute the error image. From left to the right, columns are the reconstructed image with CoSch-CLEAN, MS-CLEAN, and CS. From top to bottom, the rows present the reconstructed images in Jy/beam, the model images in Jy/pixel, and the residual image.

\begin{figure*}[!ht]
  \centering
  \begin{ocg}{fig:CygVLA}{fig:CygVLA}{0}%
    \includegraphics[width=\hsize]{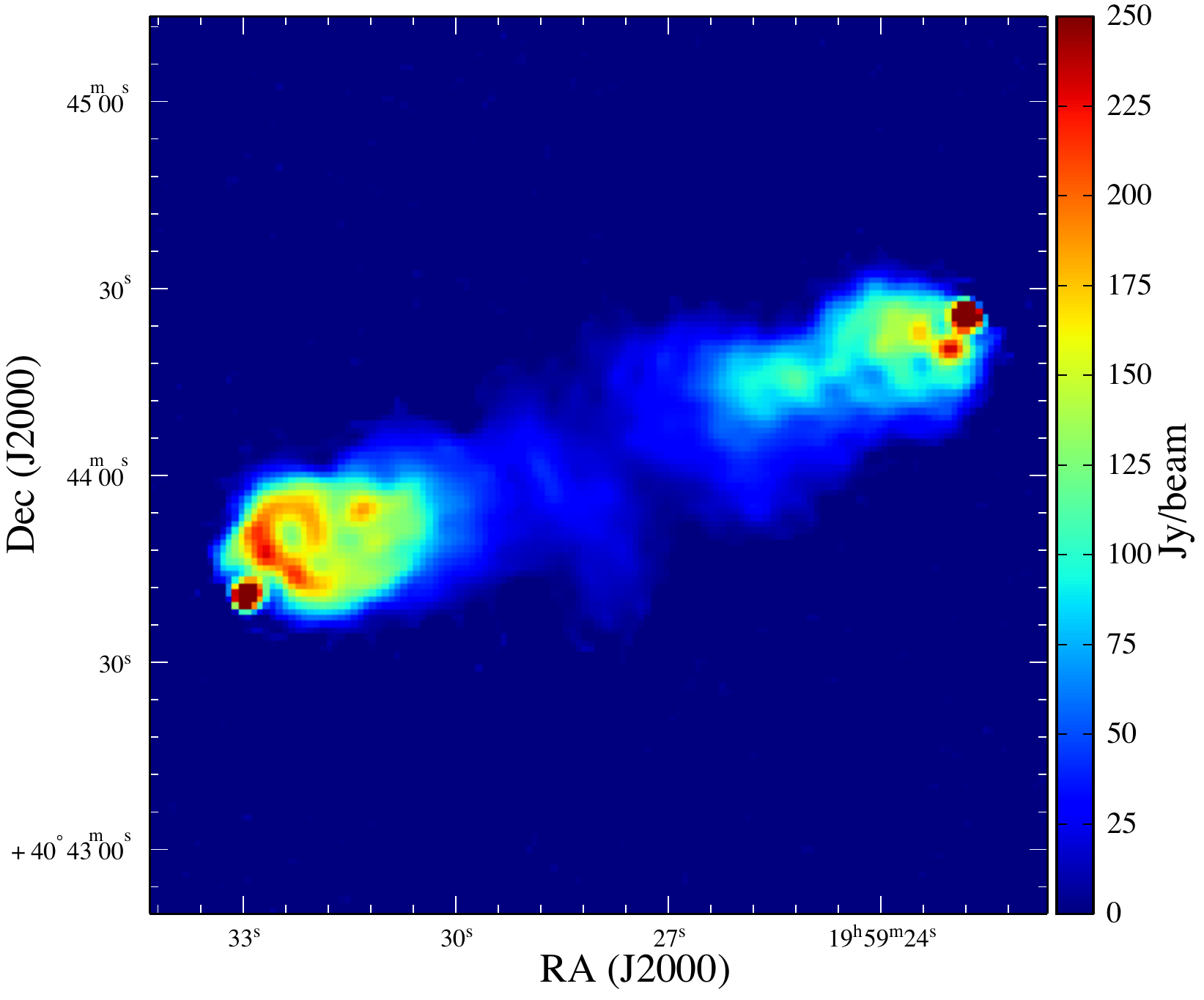}%
  \end{ocg}%
  \hspace{-\hsize}%
  \begin{ocg}{fig:CygVLAc}{fig:CygVLAc}{1}%
    \includegraphics[width=\hsize]{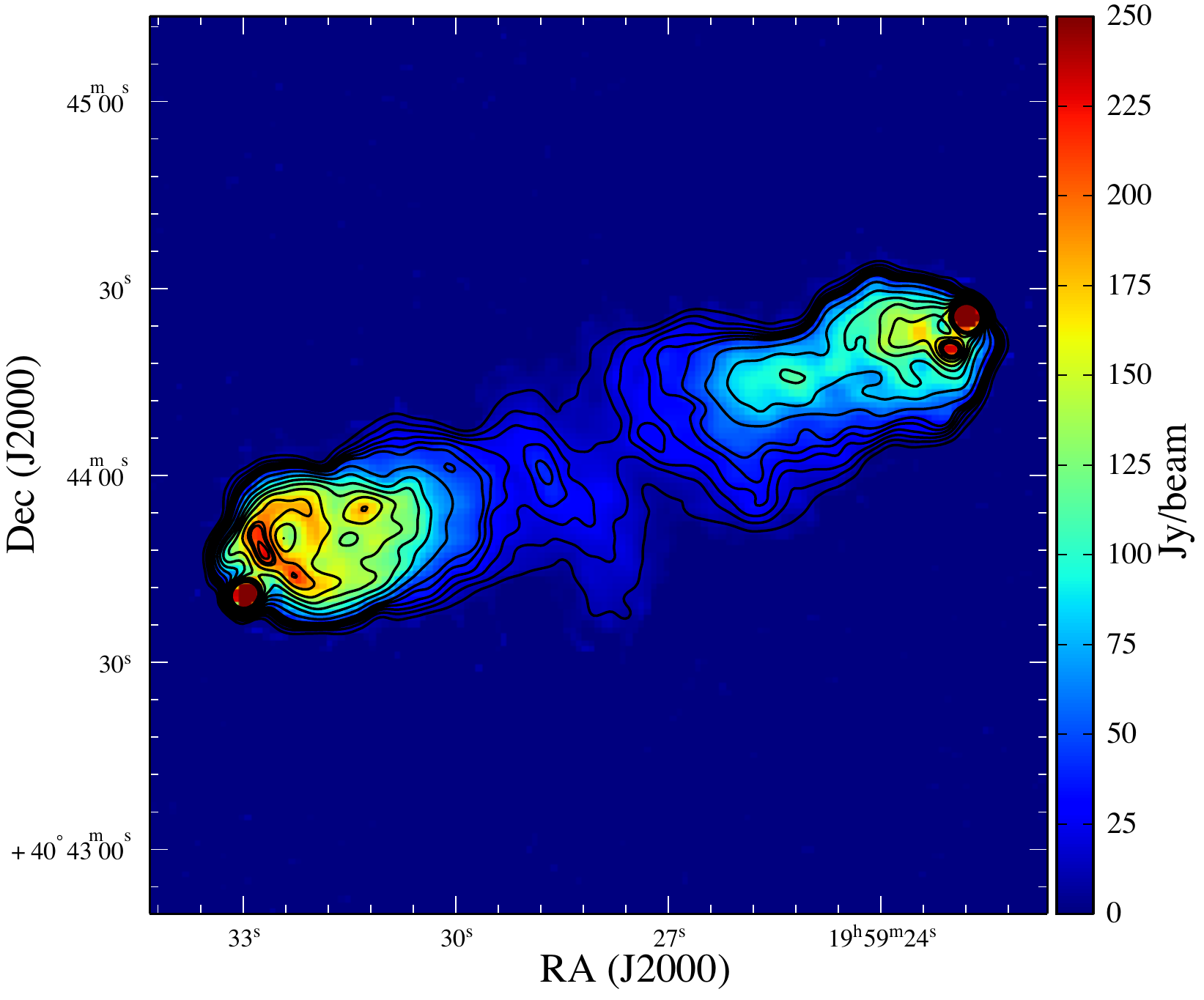}%
  \end{ocg}
\caption{Colour scale: reconstructed 512$\times$512 image of Cygnus A at 151 MHz (with resolution 2.8$''$ and a pixel size of 1$''$). Contour levels (\ToggleLayer{fig:CygVLAc,fig:CygVLA}{\protect\cdbox{Toggle contours}}) are [1,2,3,4,5,6,9,13,17,21,25,30,35,37,40] Jy/Beam from a 327.5 MHz Cyg A VLA image (Project AK570) at 2.5$''$ angular resolution and a pixel size of 0.5$''$. Most of the recovered features in the CS image correspond to real structures observed at higher frequencies.}
  \label{cygnus_lofar_vla}
\end{figure*}

All the three methods rendered a proper image of the emission, given the extremely good quality of the data and the strong source brightness.
The CoSch-CLEAN image shows residual high-frequency structures that lead to distortions of the source. These artefacts mainly come from CLEAN components collected from the noise background residual image. With MS-CLEAN, we obtained a similar brightness distribution to that in \citet{mckean10}. The restoring beam size was 6.1$''$$\times$5.0$''$ with both x-CLEAN methods. The CS effective angular resolution was $\sim$2.8$''$$\times$2.8$''$, representing again an improvement of a factor of 2. The resulting CS image shows a much sharper reconstruction of the structures inside the lobes. In particular, hotspots A, B, and D were correctly rendered without any ambiguity compared to the result obtained with both CoSch-CLEAN and MS-CLEAN. To validate the reality of the features in the CS image, we overlaid contours from VLA data at 327.5 MHz (resolution of 2.5$''$) in Fig. \ref{cygnus_lofar_vla}. The location of hotspots and filaments and holes inside each radio lobe match that of the VLA image. This confirms the improvement in angular resolution brought by CS.


As for W50, the CoSch-CLEAN model image is only composed of pixel CLEAN component sources, the MS-CLEAN image displays a smoother distribution of the source, based on the multi-scale decomposition and the CS (model) image is mainly dominated by the signal of the hotspots situated at each extremity of the lobes. 
We checked the scientific relevance of the reconstructed images by inspecting the figures of merit gathered in Table \ref{summarystat}. The first quantity is the total integrated flux density of the source. This value can be measured with short spacings (ideally with a 0-length baseline or with the autocorrelations of the antennas) or by fitting the expected visibility function in amplitude vs. the (u,v) radial distance and measuring the Y-intercept of this curve. In the present case, the observing frequency was 151 MHz and the total expected flux density is $\gtrsim$10500 Jy from the visibility data. A correct image reconstruction is therefore obtained when the total flux density present in the reconstructed image is close to that value.  
We found 10576 Jy, 10560 Jy, and 10507 Jy for CoSch-CLEAN, MS-CLEAN, and CS. This range of total flux densities is compatible with previous MERLIN and LOFAR observations taken at the same frequency \citep{steenbrugge2010,mckean10}, within a 2\% accuracy. We cannot conclude on the superiority of one method over another, but these values suggest again that CS conserves the total flux (as seen for W50) even with (real) noise, similarly to the two other methods.
The residual images are represented with the same colour scale in Fig. \ref{cygnus_panel}.
We measured the r.m.s level over the same region $\mathbf{S}$ of the flux density integration and the Std-dev of the entire 512$\times$512 residual image. The CS residuals show an improvement of about one order of magnitude over CoSch-CLEAN and MS-CLEAN. This result is characteristic of a reliable reconstruction.
The dynamic range (DR), computed as the ratio of the peak flux density to the residuals Std-dev, was computed for each image. The DRs are 1799:1, 1619:1, and 8392:1, suggesting that CS enhances the DR of the image. This is achieved by combining two effects: first, CS tends to concentrate the flux at the correct astrometric position, resulting in a higher peak flux density of the image; second, the low standard deviation of the residuals demonstrates a better convergence of the image reconstruction.  The DR of the MS-CLEAN image is slightly lower than that of CoSch-CLEAN. This can be explained by the fact that the data are dominated by the source, making it easier to decompose in unambiguous CLEAN components, or by an unoptimized choice of the scale decomposition. 
From the present results, it appeared that the CS method is reliable enough to reconstruct extended radio emissions coming from simulated data as well as real LOFAR data. The high flux density of Cygnus A and the quality of its calibration contributed to the quality of the CS reconstruction. 

%
%
\begin{table*}
\begin{center}
\begin{tabular}{ |c|c|c| }
\hline
  \multirow{2}{*}{{\bf Point source (Sect. \ref{angularsep})}} &  \multirow{2}{*}{{\bf CoSch-CLEAN}}  &  \multirow{2}{*}{{\bf CS}}\\ 
& & \\ \hline
Relative flux density error high S/N, low S/N & 3\% -- 23\%& 3\% -- 25\%\\ \hline
Effective angular resolution (') & 3.18$'$$\times$2.55$'$& 1.55$'$$\times$1.09$'$ \\ \hline
Astrometric error (') & $<$2$'$ & $<$1$'$\\ \hline
 \multirow{2}{*}{ {\bf Grid of point sources (Sect. \ref{WFsourcegrid})}} &  \multirow{2}{*}{{\bf CoSch-CLEAN}}  &  \multirow{2}{*}{{\bf CS}}\\
& & \\ \hline
Relative flux density error (Jy) & \multicolumn{2}{|c|}{Fig. \ref{point_sources}}   \\ \hline
Residuals: Std-dev in \textbf{I} (Jy/beam) & 1.7 &  3.6\\ \hline

 \end{tabular}
 \vspace{5mm}

\begin{tabular}{ |c|c|c|c| }
\hline
 \multirow{2}{*}{{\bf W50 (Sect. \ref{secw50})}} &  \multirow{2}{*}{{\bf CoSch-CLEAN}}  &  \multirow{2}{*}{{\bf MS-CLEAN}} &  \multirow{2}{*}{{\bf CS}}\\
& & & \\ \hline

Effective angular resolution (') & 2.8$'$$\times$2.4$'$& 2.8$'$$\times$3.3$'$ & $\sim$1$'$$\times$$\sim$1$'$ \\ \hline

Reconstructed: total flux density in \textbf{S} (Jy)&229.19 & 229.41  & 229.72\\ \hline
Reconstructed: Peak flux density in \textbf{I} (Jy/beam)& 0.72  & 0.96 & 0.79 \\ \hline
Residuals: r.m.s. in \textbf{S} (Jy/beam) & $4.6.10^{-4}$ & $2.71.10^{-2}$ & $3.1.10^{-4}$ \\ \hline
Residuals: Std-dev in \textbf{I} (Jy/beam) & $4.1.10^{-4}$   & $1.3.10^{-2}$  & $3.2.10^{-4}$  \\ \hline
Error: Mean square error in \textbf{I} (Jy/pixel)&  $8.9.10^{-5}$ & $1.0.10^{-4}$ & $4.8.10^{-5}$\\ \hline

DR & 1729:1 &  75:1 &  2446:1\\ \hline
 
  \multirow{2}{*}{{\bf Cygnus A (Sect. \ref{seccygA})}} & \multirow{2}{*}{{\bf CoSch-CLEAN}}  & \multirow{2}{*}{{\bf MS-CLEAN}} & \multirow{2}{*}{{\bf CS}}\\
 & & & \\ \hline
 Effective angular resolution (') & 6.1$''$$\times$5.0$''$&6.1$''$$\times$5.0$''$& 2.8$''$$\times$2.8$''$ \\ \hline
Reconstructed: total flux density in \textbf{S} (Jy)& 10576& 10560 & 10506\\ \hline
Reconstructed: Peak flux density in \textbf{I} (Jy/beam)& 315.5  &309.3 & 465.7 \\ \hline
Residuals: r.m.s. in \textbf{S} (Jy/beam) & 0.37 & 0.35 & 0.09 \\ \hline
Residuals: Std-dev in \textbf{I} (Jy/beam) & 0.17  & 0.19  & 0.05  \\ \hline
DR & 1799:1 & 1619:1 & 8392:1\\ \hline
 \end{tabular}

%
\caption{Statistical results obtained from applying CoSch-CLEAN, MS-CLEAN, and CS to all the datasets \textbf{A}, \textbf{B}, and \textbf{C}. The statistics are defined in the text and appear when applicable. \textbf{I} means that the quantity has been evaluated on the entire image, \textbf{S} in a defined region of the image (typically around the source).}
 \label{summarystat}
\end{center}
\end{table*}

\section{Discussion}
\label{discussion}

\subsection{Instrumental limitations and final convolution}
\label{convolution}
The final step of CLEAN includes convolving the model image with the CLEAN beam in the pixel domain. Each detected point source finally has the shape of the CLEAN beam, resulting in a voluntary degradation of the information included in the model image. This is a way to express the ``natural'' angular resolution of the instrument, imposing a boundary on the size of the smallest ``relevant'' feature in the image. 
The choice of the final convolution beam (or any other beam) results also from an aesthetic rendering at the expense of realness.
The CS output image is assumed to be (or at least close to) the true sky under certain conditions (see Sect. \ref{descCS}). We demonstrated that it was possible to push the limit of angular resolution. The resulting CS model image does not have an infinite resolution (it would require an infinite Fourier support) and the equivalent CLEAN beam size is limited by the quality of the Fourier reconstruction at high spatial frequencies. Preliminary studies suggested that this size may improve with number of iterations.
In addition, we chose not to convolve the final CS result with the CLEAN beam to preserve the recovered super-resolved features. As a consequence, the flux density of point sources is concentrated in smaller patches around the sources, resulting in higher pixel flux density values on Jy/beam images obtained with CS (and scaled with the CLEAN beam area).

\subsection{Performance and convergence compared to CLEAN-based methods}
We conducted several preliminary tests, which led to an educated guess of each CS parameter value used in this study.
CoSch-CLEAN and CS only differ by the core of their minimization algorithm (iterative subtraction and iterative soft-thresholding). The run time of the imager is dominated by the time spent with operators $\mathbf{A}$ and $\mathbf{A^T}$, used to convert \textit{gridded} image data to \textit{ungridded} visibilities (and vice versa) in the context the of RIME. As an example, with the W50 dataset, a single iteration of CoSch-CLEAN in \textit{AWimager}) takes 1.88s to run where CS takes 1.76s.  The number of iterations can vary significantly between CoSch-CLEAN, MS-CLEAN and CS and leads to different convergence rates that are not easily comparable. It tends to be more stable for CS because the latter always operates on the entire image during a single cycle.
It appeared that the MAD estimation (Eq. \ref{MAD}) is particularly reliable in the presence of a high level of noise, but produces
similar results to fixed thresholding when the signal of the source is dominant in the data.



\section{Conclusions and perspectives}
\subsection{Conclusion on the benchmark}
The overall performance of CS, highlighted in this study, is very promising in the scope of LOFAR, and more generally in classical radio interferometry imaging. Throughout our studies, we can draw conclusions on the behaviour of CS:

\begin{itemize}

\item CS performs relatively well compared to CLEAN in almost all situations. It was able to correctly recover the positions and flux densities of single point sources while bringing an improved angular resolution at high and moderate S/N regimes, breaking the limit imposed by the chosen CLEAN beam (or more widely by the instrumental PSF). At low S/N regimes (S/N $\lesssim$ 5), the angular separability of the CS applied to point sources falls back to that of CoSch-CLEAN.

\item CS is compatible with the RIME framework (thus enabling the correction for ionospheric effects, complex and varying antenna beams, pointing errors, etc.) and it provides satisfying photometry when imaging wide FoV. Both CS and CoSch-CLEAN had the same difficulty in recovering the lowest flux density sources. The dynamic range might be improved by a correct fine-tuning of the imaging parameters with the different methods. A more sophisticated way of studying point flux density recovery would be to simulate many observations, each with a single point in the FoV, and at different random locations. 

\item CS is able to reconstruct extended emission with improved angular resolution. From the simulated dataset (W50) and the real LOFAR dataset (Cygnus A), CS has demonstrated its ability to recover unsampled regions of the visibility function, leading to the rendering of realistic features at higher spatial frequencies than those of the two other methods. The latter are more adapted to sparse distributions of point sources than to extended continuous sources, as depicted by the vestiges of sources that remain in both error and residual images. But even if CoSch-CLEAN is limited when applied to extended emission, MS-CLEAN is still a very good alternative to reduce the image distortion and improve the sparsity as compared to CoSch-CLEAN.

\item We very carefully integrated our code in the existing LOFAR tools, that is, made it compatible with an immediate and a standard use with LOFAR. In particular, our current implementation of CS exists within \textit{AWimager} and \textit{LWimager}, which is the standard LOFAR imager that can handle DDE (by performing both \textit{A-} and \textit{W-projection}) as well as real data.
\end{itemize}

\subsection{Future developments}
In the CS algorithm, the -- iterative soft thresholding -- we used is not sophisticated, and other more sophisticated algorithms exist, such as SARA \citep{carrillo12,Carrillo2014}. There is much scope for modifying and experimenting with different minimization algorithms and their parameters, and this will be continued in the future and reported in future publications. In particular, we will integrate a reweighted-L1 scheme, which was demonstrated to reduce the bias of the signal reconstruction \cite{candesR1}.

Performance improvements will be made to the computation of $\mathbf{A}$, for example, by using multi-threading and GPGPUs \citep{hardy13}. This will benefit both CoSch-CLEAN and CS. There is a possibility of splitting CS into a major and minor cycle, similarly to CoSch-CLEAN, which will improve its speed.
For our current CS implementation, a rigorous convergence criterion should be defined. The cycle stops either when reaching $N_{max}$ or if the improvement between two consecutive iterations falls below a predefined threshold $\epsilon$ (Eq. \ref{minimize_l0_pix}). We set this number to 200, as derived from educated experimentation on the presented datasets. Future implementation will include the tracking of the convergence based on the residual r.m.s., and the determination of a robust stopping criterion.

The previous results suggest that CS methods are now mature enough for wider applications involving real datasets from LOFAR or other radio interferometers. To our knowledge, considering other completed developments or those that are in preparation as well
as those discussed in the literature, this is the first time that CS has been integrated into an existing imaging system that handles DDEs through \textit{A} and \textit{W-projection} and operates on real data produced by a current giant radio telescope. To generalize it to other radio interferometers, we are now continuing this work by integrating our code into the standard imager of the CASA (\textit{common astronomy software applications}) NRAO software package (\textit{casapy}).

We plan to study the reliability of CS when applied to different visibility data sampling (i.e. sparse measurements brought by snapshot observations vs. non-sparse measurements of long time-integrated observations), different DDEs occurring with radio interferometers such as LOFAR, and various types of (resolved) objects with a high dynamic range, and requiring different dictionaries (e.g. imaging of other complex and extended objects at low frequencies such as Virgo A, which is composed of a mix of extended emission and point sources). Reconstruction with a curvelet dictionary \citep{curvelets} is also possible in our implementation and leads, for the moment, to similar results. This specific dictionary further improves the sparsity of the solution when the source contains filamentous structures. Its extensive use will be a subject of future CS investigations in radio interferometry. 

Future applications of CS will address multi-frequency imaging and imaging of transient sources that change during the course of an observation, and hence are imprinted only partially on the visibilities. CS is expected to provide the capability to recover snapshot images from sparsely sampled uv-coverage data slices. Multi-frequency polarization imaging can also be an application of this extension at low frequencies, see \cite{Li2011RMS}.

Another aspect would be to investigate the use of CS for VLBI (\textit{very long baseline interferometry}) imaging in LOFAR. By construction, an interferometer such as LOFAR presents a statistically lower number of very long baselines than the high density of shorter baselines. The highest angular resolution of the image depends on the quality of the measurement performed at these baselines. The signal from the longest baselines is sparse by nature, noisy, and especially sullied with DDEs (e.g. the radio antennas are situated under different ionospheres). It is therefore of interest to find methods such as CS that can reconstruct the true signal from these baselines.

Currently, the next generation of radiotelescopes are developed, namely SKA. The computational effort required to perform the observation, data acquisition, distribution, and processing calls for developing hierarchical and distributed real-time processing. These developments rely on the success of the current SKA \textit{pathfinders} and \textit{precursors}\footnote{https://www.skatelescope.org/technology/precursors-pathfinders-design-studies}. 
In that context, CS can be used for instrumental calibration. As an example, the \textit{Murchison Wide-field Array} (MWA) has routinely deployed a fast-imaging mode to perform real-time image-plane calibration \citep{mwa2008}. Such a calibration mode, combined with CS methods that can approximate the content of the true sky from very sparse measurements, can represent a major advance in instrumental calibration. The NenuFAR project \citep{nenufar2012}, the independent phased-array interferometer formed from the extension of the French LOFAR station \textit{FR606}, will have similar requirements. Moreover, this local interferometer offers a densely populated uv-space below $B$$\sim$400 m with rare E-W and N-S longer baselines, which may benefit from CS reconstruction methods.

A larger community of astronomers (and especially imagers) are using and confirming the relevance of CS reconstruction methods. In the context of radio interferometry, we may use the sparse reconstruction method as a new standard way to image or calibrate data, complementary to x-CLEAN algorithms. The coming generation of radio instruments will be resource-demanding and will provide many potential applications for CS.

\subsection{Software availability}
\label{soft}
An independent version of the CS software is available\footnote{http://www.cosmostat.org/software.html} for testing the CS code with different toy models. This package is called SASIR (\textit{sparse aperture synthesis interferometry reconstruction}) and comes with its source code. It demonstrates the feasibility of the sparse reconstruction using either an original model image and a Fourier mask in FITS format (representing the (u,v) sampling of the instrument), or a complex FITS image coming from gridding the data from any given measurement set. The operator $\mathbf{A}$, central to the experiments of this paper, is computed directly inside the \textit{AWimager} and cannot be extracted explicitly from the LOFAR framework, as it changes for every measurement set and refers to specific LOFAR dependencies for computing the DDE (antenna beam, polarization, projection effects, etc.). The source code of \textit{AWimager} can be obtained by contacting Cyril Tasse and colleagues \citep{tasse13}.

\begin{acknowledgements}
HG and JNG were responsible for the implementation of the CS code in the LOFAR imager (AWimager), JNG wrote most of the article and performed all the scientific validation on simulated and real data. JLS and SC conducted the research, supplied support, CS libraries, and wrote the independent version (SASIR) of the code. CT participated to the implementation of AWimager for LOFAR. AW wrote the preliminary version of the independent code. JPM provided the LOFAR Cygnus A dataset. The other co-authors are part of the LOFAR builders list. We would like to thank all co-authors and the anonymous referee for their active contribution and/or insightful comments and reviews.
We acknowledge the financial support from the UnivEarthS Labex program of Sorbonne Paris Cit\'e (ANR-10-LABX-0023 and ANR-11-IDEX-0005-02) and from the European Research Council grant SparseAstro (ERC-228261).
LOFAR, the Low Frequency Array designed and constructed by ASTRON, has facilities in several countries, that are owned by various parties (each with their own funding sources), and that are collectively operated by the International LOFAR Telescope (ILT) foundation under a joint scientific policy.
Chiara Ferrari acknowledges financial support by the {\it Agence Nationale de la Recherche} through grant ANR-09-JCJC-0001-01.
VLA data of 3C 405 come from the online NRAO Archive (the National Radio Astronomy Observatory is a facility of the National Science Foundation operated under cooperative agreement by Associated Universities, Inc.).  
\end{acknowledgements}

\bibliographystyle{aa} 
\bibliography{refs} 

\end{document}